\documentclass[10pt,journal]{IEEEtran}
\IEEEoverridecommandlockouts

\usepackage{graphicx}
\usepackage{subfigure}
\usepackage{amsmath,amsthm,amsfonts,amssymb}
\usepackage{cite}
\usepackage{bm}
\usepackage{bbm}
\usepackage{url}
\usepackage{array}
\usepackage{color}
\usepackage{multirow}
\usepackage{booktabs}
\usepackage[table,xcdraw]{xcolor}
\usepackage{enumerate}

\theoremstyle{plain}
\newtheorem{thm}{Theorem}
\newtheorem{lem}[thm]{Lemma}

\newtheorem{prop}[thm]{Proposition}

\newtheorem*{rem}{Remark}
\newtheorem{sty1}{Theorem}
\newtheorem{defi}[sty1]{Definition}

\newenvironment{NewProof}{{\noindent\it Proof.}}{\hfill $\blacksquare$\par}

\usepackage[english]{babel}
\usepackage{algorithm}
\usepackage[noend]{algpseudocode}

\allowdisplaybreaks[4]

\begin{document}
\title{Flow Sampling: Network Monitoring in Large-Scale Software-Defined IoT Networks}

\author{
Yulin~Shao,~\IEEEmembership{Member,~IEEE},
Soung~Chang~Liew,~\IEEEmembership{Fellow,~IEEE},
He~Chen,~\IEEEmembership{Member,~IEEE},
Yuyang~Du
\thanks{The authors are with the Department of Information Engineering, The Chinese University of Hong Kong, Shatin, New Territories, Hong Kong (e-mail: \{sy016, soung, he.chen, dy020\}@ie.cuhk.edu.hk).}
}

\maketitle
\begin{abstract}
Software-defined Internet-of-Things networking (S\-DIoT) greatly simplifies the network mo\-nitoring in large-scale IoT networks by per-flow sampling, wherein the controller keeps track of all the active flows in the network and samples the IoT devices on each flow path to collect real-time flow statistics.
There is a tradeoff between the controller's sampling preference and the balancing of loads among devices.
On the one hand, the controller may prefer to sample some of the IoT devices on the flow path because they yield more accurate flow statistics.
On the other hand, it is desirable to sample the devices uniformly so that their energy consumptions and lifespan are balanced.
This paper formulates the flow sampling problem in large-scale SDIoT networks by means of a Markov decision process and devises policies that strike a good balance between these two goals.
Three classes of policies are investigated: the optimal policy, the state-independent policies, and the index policies (including the Whittle index and a second-order index policies).
The second-order index policy is the most desired policy among all:
1) in terms of performance, it is on an equal footing with the Whittle index policy, and outperforms the state-independent policies by much;
2) in terms of complexity, it is much simpler than the optimal policy, and is comparable to state-independent policies and the Whittle index policy;
3) in terms of realizability, it requires no prior information on the network dynamics, hence is much easier to implement in practice.
\end{abstract}


\section{Introduction}
The scale of Internet-of-Things (IoT) networks is getting increasingly large in recent years. According to Cisco \cite{Cisco}, the number of connected IoT devices is due to reach $75$ billion by $2025$. In the foreseeable future, massive IoT networks with millions of connections can pose great challenges to network monitoring, management, and control \cite{SDIoT1,challenge2007,APNC}.

To enable efficient and dynamic network monitoring, a recent trend in large-scale IoT networks is the fusion of software-defined networking (SDN) \cite{SDNMagazine} and IoT, dubbed software-defined IoT networking (SDIoT)~\cite{SDIoT1,SDIoT2,TCOM,Significant2020}. The essence of SDIoT, as that of SDN, is to disassociate the data plane (forwarding process of network packets) from the control plane (monitoring and configuration). In particular, the control plane is managed by one or more logically centralized controllers that have a global view of the network. SDIoT greatly simplifies the network monitoring process in large-scale IoT networks because all the IoT devices are equipped with programmable interfaces \cite{monitoring}, with which the controller can query/sample the devices for statistics of each flow passing through them. In a nutshell, the controller monitors the SDIoT network by simple means of per-flow sampling \cite{openTM,Opennetmon,DCM,SLAM}.

As an example, let us elaborate on OpenTM \cite{openTM}, a network monitoring system implemented on SDN, to explain how per-flow sampling works in large-scale networks. The goal of OpenTM is to estimate the traffic matrix (TM) of the network, i.e., a traffic map reflecting the volumes of traffic flowing between all the edge devices of the network \cite{TMSigMetrices,TMprimer}. To this end, OpenTM keeps track of the statistics of all active flows in the network and monitors each flow independently. For each active flow, the controller 1) gets the routing information and determines the flow path; 2) periodically samples flow statistics, such as flow byte and packet count, from one of the devices on the flow path; 3) constructs the TM by adding up statistics of flows from the same source to the same destination.

As can be seen, per-flow sampling is well suited for large-scale SDIoT networks in that
1) The controller can directly communicate with the IoT devices thanks to the measurement infrastructure provided by the SDIoT. This allows lightweight sampling operations that yield real-time flow statistics.
2) In lieu of centralized sampling, per-flow sampling scales well with the network size and adapts to the dynamic nature of the IoT-network topology -- IoT devices are often deployed without paying particular attention to the topology they form and the post-deployment topology can change frequently because of the displacement of IoT devices, e.g., industrial IoT (IIoT) networks.

Consider sampling a single flow path. A decision to be made by the controller is which IoT device to sample in each sampling epoch. In traditional SDN, an important criterion to be considered when making the sampling decision is the sampling preference of the controller \cite{openTM,Opennetmon,SLAM,FlexMonitor}.

\textbf{Sampling preference of the controller} -- The controller may have a preference to sample some of the devices on the flow path. In OpenTM, for example, different devices on a flow path can observe different traffic volumes for the flow due to the packet loss. OpenTM aims to capture the amount of traffic that arrives at the destination. Therefore, the controller prefers to sample the last device of the path \cite{openTM} because it is closest to the destination, and the traffic volumes sampled from it are considered as the most accurate.

The sampling preference of the controller also exists in other applications. OpenNetMon \cite{Opennetmon} and SLAM \cite{SLAM} are OpenFlow controller modules developed to measure per-flow latency, packet loss, and throughput. For these intents, the controller prefers to sample the first and the last devices because the difference between the statistics collected from them gives the most accurate measurements. In FlexMonitor \cite{FlexMonitor}, on the other hand, the controller prefers to sample the devices that yield the minimal communication cost with the controller.

In addition to the sampling preference of the controller, another design dimension that merits particular treatment in SDIoT networks is the load balancing among IoT devices.

\textbf{Load balancing among IoT devices} -- Devices consume extra energy to execute the sampling tasks. However, unlike the traditional SDN wherein the network nodes are routers and switches that are connected to power supplies, the network nodes in SDIoT are low-powered IoT devices. Therefore, a fair sampling policy should be able to distribute the sampling tasks evenly to the IoT devices so that the energy consumptions and lifespan of different devices on the flow path are balanced \cite{DCM,Tradeoff2013}.

There is a clear tradeoff between the above two criteria. As far as the sampling preference is concerned, the controller prefers to sample some of the IoT devices more frequently as they yield more accurate flow statistics. On the other hand, in terms of load balancing, it is preferred to sample the IoT devices uniformly so that they carry equal average loads. An outstanding issue in SDIoT networks is how to devise a judicious flow-sampling policy that strikes the best tradeoff between these two criteria.

To fill this gap, this paper formulates the flow sampling problem in SDIoT networks and investigates different sampling policies that balance the controller's sampling preference (i.e., more accurate statistics) and load balancing among IoT devices. In particular, we model the flow sampling problem as a discrete Markov decision process (MDP) \cite{MDPBook,POMDP} with the {\it state} being a measurement of load balance among devices.
The sampling policy of the controller is a mapping from state to an {\it action} (i.e., a chosen device), and different actions yield different sampling accuracies.
In successive time slots, the controller follows its sampling policy and makes a sequence of independent decisions to sample one of the IoT devices on the flow path.
The quality of an action at a state is reflected by a {\it cost} associated with this state-action pair. This cost function is designed to take both sampling accuracy and load balancing among IoT devices into account. The optimal sampling policy is then defined as the policy that minimizes the average cost on an infinite time horizon.

Three classes of sampling policies are explored in this paper as solutions to the MDP: the optimal policy, the state-independent policies, and the index policies.

\textbf{The optimal policy} -- The optimal policy is derived by solving the MDP using stochastic dynamic programming (DP) \cite{MDPBook}. Although optimal, the relative value iteration algorithm for stochastic DP is computationally intensive: its complexity grows exponentially with the increase of the number of IoT devices on the flow path. This limits the scalability of stochastic DP when the sampling problem involves a large number of IoT devices.

\textbf{State-independent policy} -- As the name suggests, state-independent policies make the sampling decision without considering the current states of the IoT devices. We analyze two state-independent policies implemented in OpenTM \cite{openTM}: a uniform sampling policy and a non-uniform sampling policy. For each flow path, the uniform policy instructs the controller to sample the IoT devices uniformly at random. The non-uniform policy, on the other hand, indexes the devices on the flow path so that devices with larger indexes are closer to the destination. In each decision epoch, the non-uniform policy randomly generates two integers and instructs the controller to sample the device indexed by the larger integer. We further generalize these two state-independent policies to a largest-order-statistic policy and a weighted-probability policy that have better performance. In particular, the weighted-probability policy is the optimal stationary state-independent policy.

Overall, state-independent policies have very low complexity, hence are easy to implement in practice. Their performance, however, is suboptimal in general.

\textbf{Index policies} -- To devise low-complexity policies with good performance, we consider a class of index policies to solve the MDP. The Whittle index \cite{Whittle1988} refers to an index policy proposed by Whittle to solve restless multi-armed bandit (RMAB) problems \cite{GittinsBook}. RMAB is a sequential decision problem where, at each time, one or more choices must be made among all available Markovian arms/jobs. The Whittle index associates each arm with an index, and chooses the arm with the largest index at each decision epoch \cite{Whittle1988}. By so doing, the original high-dimensional decision problem is decoupled to multiple one-dimensional problems of computing the individual indexes of the jobs/arms, hence the computational complexity of the Whittle index grows linearly in the number of arms. Thanks to its low complexity and excellent performance, the framework of RMAB and the Whittle index solution has been widely used to solve the problem of route planning for unmanned military aircraft \cite{RMABUAV}, opportunistic communication channel usage \cite{RMABIT,Kadota1}, and sensor management \cite{RMABSensor}, to name a few.

This paper formulates our MDP as an RMAB problem and devises a Whittle index policy to solve the MDP. The Whittle index is derived in closed form. Simulation results show that 1) the Whittle index policy performs as well as the optimal policy derived from stochastic DP when the number of IoT devices on the flow path is small; 2) the Whittle index policy outperforms all the state-independent policies. Compared with the uniform policy and the largest-order-statistic policy, the Whittle index policy reduces the average cost by $66.4\%$. Compared with the weighted-probability policy, the Whittle index policy reduces the average cost by $33.4\%$.

The Whittle index policy has satisfactory average-cost performance and low computation complexity. Yet, as the optimal policy does, it relies on perfect knowledge of the network dynamics for ``planning''. This prior knowledge, however, may not be available to the controller in practice. In view of this, this paper further puts forth a second-order index policy inspired by the form of the Whittle index. The second-order index policy is the most desired policy among all as it requires no prior knowledge of the network dynamics while having all the advantages of the Whittle index. Simulation results show that the performance gap between the second-order index policy and the Whittle index is negligible.

\section{Problem Formulation}\label{sec:II}
\begin{figure}[t]
  \centering
  \includegraphics[width=0.7\columnwidth]{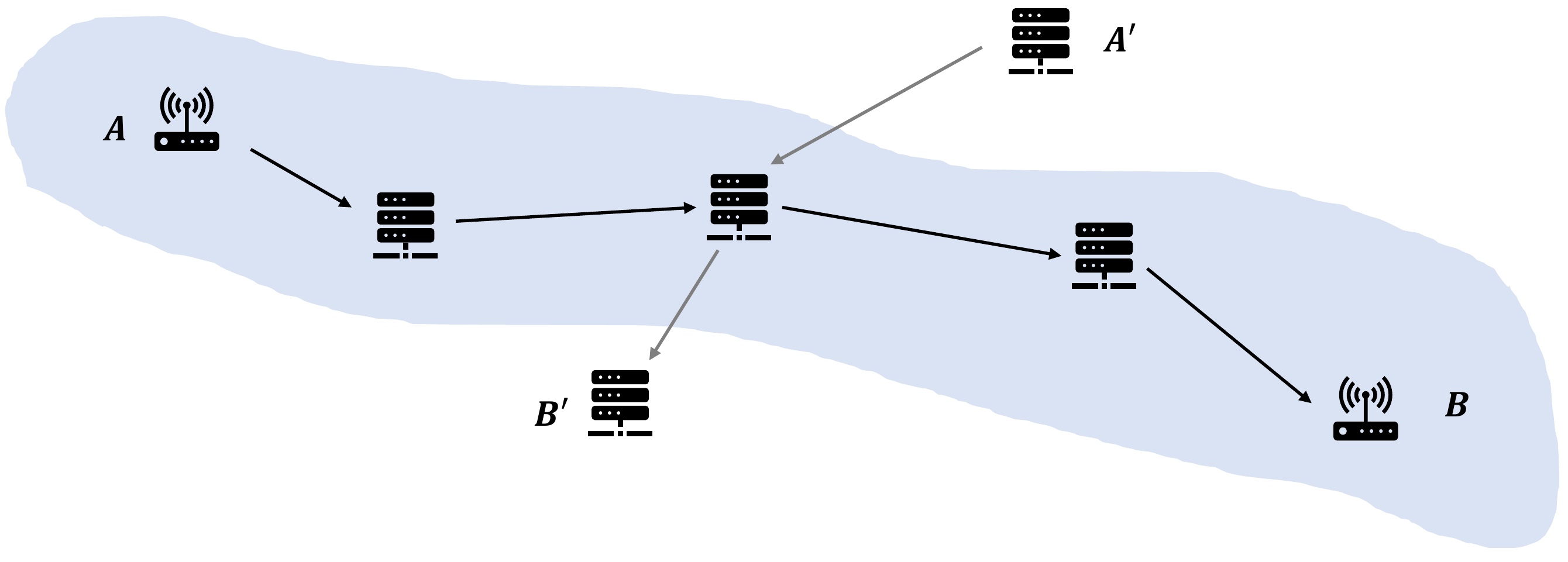}\\
  \caption{A flow path $\overline{AB}$ from origin A to Destination B with $M = 3$ IoT devices. Another flow path $\overline{A'B'}$ intersects with $\overline{AB}$ at the second device.}
\label{fig:1}
\end{figure}

Consider an SDIoT network where the controller monitors all active flows in the network. In particular, the controller monitors each flow independently in a time-slotted manner,\footnote{It is worth noting that per-flow sampling is not optimal in the sense that the sampling of all active flow paths in the network are not jointly optimized.
Nevertheless, it is a scalable and flexible solution that is easy to be implemented in large-scale IoT networks considering the complex network topologies and dynamic configurations of flow paths in practice \cite{SDIoT1,SDIoT3,Opennetmon}.} and the monitoring slot boundaries for different flows can be misaligned. Without loss of generality, we focus on a single flow path $\overline{AB}$ from origin A to destination B.
As shown in Fig.~\ref{fig:1}, $\overline{AB}$ travels through $M$ IoT devices indexed by $\{i: i = 1, 2, ..., M\}$, and devices with larger indexes being closer to the destination. At the beginning of time slots $\{t: t = 0, 1, 2, ...\}$, the controller has to decide which of the $M$ IoT devices to sample to collect real-time flow statistics.

A judicious sampling policy should be devised to balance the tradeoff between sampling accuracy and fair loading among IoT devices.
Take OpenTM \cite{openTM} for instance.
Due to packet loss, the most accurate statistics of the traffic volumes arrive at the destination can be obtained from the $M$-th IoT device because it is closest to the destination. Sampling the last device in successive slots gives the controller the most accurate statistics but imposes a substantial load on the last device at the same time. If we deterministically sample the last device on each flow, all the ingress/egress IoT devices at the edge of the network will be heavily burdened with the querying loads.

\subsection{Flow Sampling}
To formulate the flow sampling problem, let us first introduce the definitions of sampling accuracy and the measurement of querying loads.
\begin{defi}[Accuracy]
We denote by $\varphi_i\in[0,1]$ the accuracy of the statistics collected from the $i$-th IoT device. The parameters $\{\varphi_i:i=1,2,...,M\}$ can take any form in general and larger $\varphi_i$ is desired in each sampling operation.
\end{defi}

To get intuitive results (and be able to compare with prior works), sometimes we may set $\varphi_i=\sigma^{M-i}$, where $\sigma\in(0,1]$ is a constant. That is, we consider a homogeneous network where the packet loss rates at the IoT devices are the same and the statistics collected from the devices closer to the destination are more accurate.

To measure the querying loads imposed on each IoT device, we let the controller maintain $M$ counters, each of which is associated with an IoT device.

\begin{defi}[Counters]
The $i$-th counter $n_i$ associated with the $i$-th IoT device records the number of slots since the last slot the $i$-th device was sampled. Over time, $n_i$ evolves in the following way:
\begin{eqnarray}\label{eq:II1}
n_i^{t+1}=\left\{
\begin{array}{lll}
0,       &&\!\!\!\!\!\!\!\!\!\!\!\!\!\!\!\! \textup{if the $i$-th device is sampled in slot $t$;}\\
n_i^t+1,     &&\!\!\!\!\!\!\!\! \textup{otherwise,}
\end{array} \right.
\end{eqnarray}
where we use superscript to denote time and subscript to denote the index of the IoT device. The counters are updated at the end of a time slot.
\end{defi}

We emphasize that the evolution of the counters in \eqref{eq:II1} can be triggered by not only the sampling of the controller on path $\overline{AB}$, but also the sampling operation on any other flow path which intersects with path $\overline{AB}$. Take Fig.~\ref{fig:1} for example. There are $M = 3$ IoT devices on path $\overline{AB}$, and there is another flow path $\overline{A'B'}$ that intersects with $\overline{AB}$ at the second device (i.e., the second device is a crosspoint). Suppose the counter array of the three IoT devices on $\overline{AB}$ are updated to $\{2,3,1\}$ at the end of slot $t-1$, and the controller decides to sample the third device of $\overline{AB}$ in slot $t$,
\begin{enumerate}[a)]
\item If the controller also samples path $\overline{A'B'}$ at the crosspoint, the counter array associated with $\overline{AB}$ evolves to $\{3,0,0\}$ because both the second and third devices are sampled by the controller in slot $t$.
\item Otherwise, if the controller does not sample path $\overline{A'B'}$ at the crosspoint, the counter array evolves to $\{3,4,0\}$.
\end{enumerate}
Succinctly speaking, the counter of an IoT device will be reset to $0$ as long as it is sampled during slot $t$, whether it is sampled by flow path $\overline{AB}$ or by any other flow paths.

In large-scale networks, the controller monitors and samples the flow paths independently.
Consider a specific path $\overline{AB}$ with $M$ IoT devices. We model the event that a device is sampled by other flow paths (other than $\overline{AB}$) as a random variable. Specifically, define the event $H^t_i$: the $i$-th device is sampled by flows other than $\overline{AB}$ in slot $t$.
We assume $H^t_i$, $\forall~i$ follows independent Bernoulli distribution with parameter $p_i$, and is time-invariant (constant over time). That is, the $i$-th device is sampled by flows other than $\overline{AB}$ with probability $p_i$ in a time slot. In this context, the evolution of $n^t_i$ in \eqref{eq:II1} can be rewritten as follows:
\begin{enumerate}[a)]
\item  If the $i$-th device is sampled by $\overline{AB}$ in slot $t$.
\begin{eqnarray}\label{eq:II_tran1}
n^{t+1}_i = 0, \textup{w. p. $1$};
\end{eqnarray}
\item If the $i$-th device is not sampled by $\overline{AB}$ in slot $t$.
\begin{eqnarray}\label{eq:II_tran2}
n_i^{t+1}=\left\{
\begin{array}{lll}
0,       &&\!\!\!\!\!\!\!\! \textup{w. p. $p_i$,}\\
n_i^t+1,     &&\!\!\!\!\!\!\!\! \textup{w. p. $1-p_i$.}
\end{array} \right.
\end{eqnarray}
\end{enumerate}

The goal of flow sampling is to discover a sampling policy that strikes good tradeoff between sampling accuracy and load balancing among devices.
An example of such a policy is the non-uniform sampling policy proposed and implemented in \cite{openTM}.
For each flow path, the non-uniform policy randomly generates two integers in each decision epoch and instructs the controller to sample the device indexed by the larger integer.
We will analyze the non-uniform policy in Section~\ref{sec:IV} and further generalize it to a largest-order-statistic policy.
In this section, let us first formulate the flow sampling problem as an MDP and define quantitatively what is a good sampling policy.

\begin{rem}
In existing implementations of flow sampling, the sampling and monitoring functions are defined in the network layer \cite{openTM,Opennetmon,SLAM,FlexMonitor}. Therefore, this paper formulates the flow sampling problem for SDIoT networks assuming error-free sampling operations thanks to the error correction code in the PHY layer and the automatic repeat request in the MAC layer.
The system model can be further generalized to a cross-layer design wherein the sampling function is defined in the PHY layer and hence is error-prone.
\end{rem}

\subsection{An MDP Formulation}
The problem of discovering the optimal flow sampling policy can be described as a discrete MDP. Specifically, at the beginning of a slot $t$, the controller observes a state of the counter array $s^t=\{n^t_i:i=1,2,...,M\}$.
Given this observation, the controller chooses an action $a^t$ (i.e., which device to sample) following its sampling policy $\mu$, and executes $a^t$ in slot $t$. The action produces two results: 1) an immediate cost $C(s^t)$ is incurred (defined later), and 2) the system evolves to a new state $s^{t+1}$ in the next slot as per the transition probability defined below.
\begin{eqnarray}\label{eq:III_tranPr}
&&\hspace{-1cm} P(s^{t+1}\left.\right| s^t, a^t=j)= \\
&&\hspace{-1cm} \qquad \prod_{i=1,2,...,M, i\neq j}\left\{p_i\mathbbm{1}_{n^{t+1}_i=0} + (1-p_i)\mathbbm{1}_{n^{t+1}_i=n^{t}_i+1} \right\},  \nonumber
\end{eqnarray}
where $\mathbbm{1}$ is an indicator function, and
\begin{eqnarray*}
&& s^t=\left(n^t_1, n^t_2, ..., n^t_{j-1}, n^t_j, n^t_{j+1}, ..., n^t_M\right), \\
&& s^{t+1}=\left(n^{t+1}_1, n^{t+1}_2, ..., n^{t+1}_{j-1}, n^{t+1}_j=0, n^{t+1}_{j+1}, ..., n^{t+1}_M\right).
\end{eqnarray*}
Eq.~\eqref{eq:III_tranPr} defines the probability that the controller evolves from $s^t$ to $s^{t+1}$ if action $a^t=j$ is executed in slot $t$. Specifically, 1) the $j$-th counter $n^{t+1}_j$ is reset to $0$ deterministically; 2) the $i$-th counter $n^{t+1}_i$, $i\neq j$ is reset to $0$ with probability $p_i$, and evolves to $n^t_i+1$ with probability $1-p_i$. The evolutions of all counters are independent. Thus, $P(s^{t+1}\left.\right| s^t, a^t=j)$ is a product of $M-1$ terms, each of which is $p_i$ or $1-p_i$, depending on the value of $n^{t+1}_i$, $i\neq j$.

The same decision problem is faced by the controller in all the subsequent slots, but with different observations and corresponding actions.

\begin{figure}[t]
  \centering
  \includegraphics[width=0.95\columnwidth]{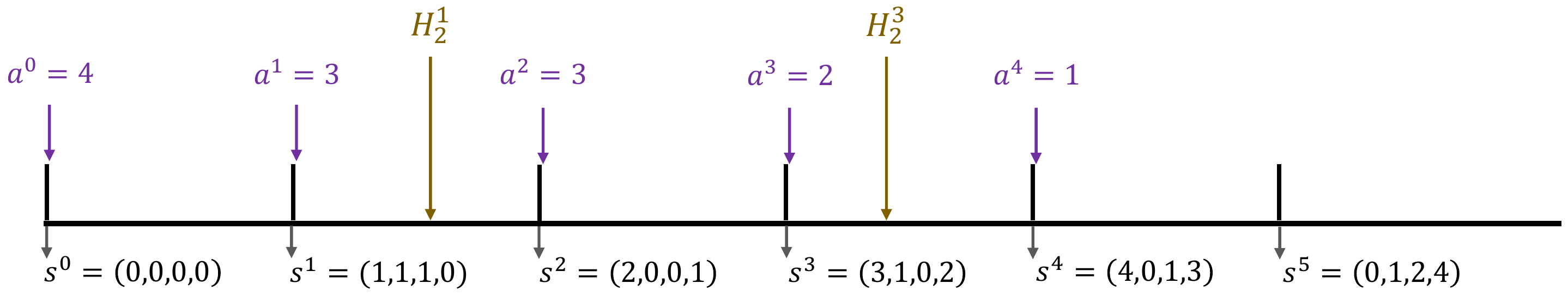}\\
  \caption{An example of the state transitions in the MDP associated with the flow sampling problem.}
\label{fig:2}
\end{figure}

An example of the state transitions is given in Fig.~\ref{fig:2}, wherein $M = 4$. As can be seen, the system starts with state $s^0=(0,0,0,0)$.
In the beginning of slot $t = 0$, the controller takes action $a^0=4$, and no event $H^0_i$ happens during slot $0$. Thus, the state transits to $s^1=(1,1,1,0)$ at the end of slot $0$ because only the fourth device is sampled.
In slot $1$, the controller takes action $a^1=3$, and there is an event $H^1_2$, meaning that the second device is a crosspoint and is sampled by another flow during slot $1$.
Thus, the state transits to $s^2=(2,0,0,1)$ at the end of slot $1$ because both the second and third devices are sampled.
In each slot, an immediate cost is incurred as the penalty of being in state $s^t$, as defined below.

\begin{defi}[Immediate cost and average cost]
The immediate cost of being in state $s^t=\{n^t_i:i=1,2,...,M\}$ is defined as
\begin{eqnarray}\label{eq:II_cost}
C(s^t)=\sum_{i=1}^{M}\varphi_i n^t_i.
\end{eqnarray}
A given policy $\mu$ instructs the controller to traverse through a series of states. The average cost incurred by this policy over the infinite-time horizon is defined as
\begin{eqnarray}\label{eq:II_avecost}
J_\mu=\lim_{T\rightarrow\infty}\mathbb{E}_\mu \left[\frac{1}{T}\sum^{T-1}_{t=0} C(s^t)   \right].
\end{eqnarray}
\end{defi}

As can be seen, we define the immediate cost to be a sum of the counter values $n^t_i$ weighted by the accuracies $\varphi_i$. In so doing, the controller favors sampling 1) the more accurate device if two or more devices have the same counter values; 2) the device with larger counter value if two or more devices are equally accurate; since it reduces the immediate cost the most.

The optimal policy, denoted by $\mu^*$, is the policy that minimizes the average cost over the infinite-time horizon, giving,
\begin{eqnarray}\label{eq:II_opt_policy}
\mu^*=\arg\min_\mu J_\mu.
\end{eqnarray}

\begin{rem} An alternative way to define the counters $n^t_i$ is the number of times that the $i$-th device has been sampled up until slot $t$. That is, counter $n^t_i$ is increased by $1$ if the $i$-th device is sampled in slot $t$, and frozen otherwise (a setup akin to the standard multi-armed bandit (MAB) problem \cite{weber1992gittins}). However, given this definition, the MDP associated with our flow sampling problem is very tricky to handle because $n^t_i$ grows indefinitely over time. In particular, the states of the MDP do not communicate. Our definition of counters in Definition \ref{eq:II1} circumvents this issue and renders the problem solvable.
\end{rem}


\section{A Lower Bound and the Optimal Policy}\label{sec:III}
To begin with, let us construct a lower bound to the average cost $J_\mu$ in \eqref{eq:II_avecost}.
This lower bound will serve as one of the benchmarks to evaluate the different policies devised in this paper.
Then, we follow classical MDP theory to derive the optimal policy in \eqref{eq:II_opt_policy}. The problems with the optimal policy are highlighted at the end of this section.

Following a sample-trajectory argument, a lower bound to the average cost $J_\mu$ can be constructed as follows.
\begin{thm}[Lower bound of the average cost]\label{thm:1}
A lower bound to $J_\mu$ is
\begin{eqnarray*}
L_B(J_\mu)=\frac{1}{2}\sum_{i=1}^M \varphi_i \left(\frac{1}{(1-p_i)\alpha^*_i+p_i} - 1 \right),
\end{eqnarray*}
where $\{\alpha^*_i:i=1,2,...,M\}$ is a distribution given by
\begin{eqnarray*}
\alpha^*_i = \left(v\sqrt{\frac{\varphi_i}{1-p_i}} -\frac{p_i}{1-p_i}  \right)^+,
\end{eqnarray*}
function $(x)^+=x$ if $x\geq 0$, and $(x)^+=0$ if $x< 0$; $v$ is chosen such that
\begin{eqnarray*}
\sum_{i=1}^{M} \left(v\sqrt{\frac{\varphi_i}{1-p_i}} -\frac{p_i}{1-p_i}  \right)^+ = 1.
\end{eqnarray*}
\end{thm}

\begin{NewProof}
See Appendix \ref{sec:AppA}.
\end{NewProof}

This lower bound is derived by assuming the variance of the inter-sampling time of each IoT device is negligible relative to the mean of the inter-sampling time.
It is supposed to be tighter in the case of larger $M$ and smaller $p_i$.

Next, we derive the optimal policy in \eqref{eq:II_opt_policy} following the classical MDP theory.
For this average-cost MDP problem, the optimal policy $\mu^*$ that minimizes the average cost on an infinite time horizon can be computed via a relative value iteration process. Specifically, the Bellman equation for the average-cost optimality criterion is given by \eqref{eq:III_bellman} below \cite{MDPBook}. We only consider stationary policies, thus the time index $t$ is removed in the rest of this section.
\begin{eqnarray}\label{eq:III_bellman}
g^*\bm{e} + \bm{h}^* = \mathcal{T}(\bm{h}^*)
\end{eqnarray}
where $g^*$ is the gain of MDP, i.e., the average cost incurred per time step when the system is in equilibrium; $\bm{e}$ is an all-ones column vector; $\bm{h}^*$ is a vector with each element being the relative value function of a state. The relative value function of a state $s$ (also named the cost-to-go function), is the difference between the total cost incurred by a system that starts with state $s$ and the total cost incurred by a system that starts with a steady-state state over an infinite time horizon, i.e., the extra cost incurred by the transient behavior of being in state $s$. The operator $\mathcal{T}$ is a Bellman operator given by
\begin{eqnarray}\label{eq:III_bellman_operator}
\mathcal{T}(\bm{h})[s] = \min_{a}\left\{C(s)+\sum_{s'}P(s'|s,a) \bm{h}[s']  \right\},
\end{eqnarray}
where $\bm{h}[s]$ is an element in the vector $\bm{h}$ that corresponds to state $s$, $P(s'|s,a)$ is defined in \eqref{eq:III_tranPr}.

\begin{algorithm}[t]
\caption{Relative value iteration}\label{algo:1}
\begin{algorithmic}[1]
\State{Pick $\bm{h_\textup{old}}$ arbitrarily, pick a reference state $s_0$ arbitrarily.}
\State{Set a stopping criterion $\epsilon$.}
\State{$\textup{SP} = \epsilon + 1$.}
\While{$\textup{SP} > \epsilon$}
\State{$\bm{h_\textup{new}}=\mathcal{T}(\bm{h_\textup{old}})-\bm{e}\mathcal{T}(\bm{h_\textup{old}})[s_0]$}
\State{$\bm{d=h_\textup{new}-h_\textup{old}}$}
\State{$\textup{SP}=\max_l \bm{d}[l]-\min_l \bm{d}[l]$}
\State{$\bm{h_\textup{old}=h_\textup{new}}$}
\EndWhile
\State \textbf{return} $\bm{h^*=h_\textup{new}}$, $g^*=\mathcal{T}\left(\bm{h_\textup{new}}\right)[s_0]$.
\end{algorithmic}
\end{algorithm}

The solution $(g^*, \bm{h}^*)$ to \eqref{eq:III_bellman} can be computed by a relative value iteration process given in Algorithm~\ref{algo:1}, the convergence of which is guaranteed as the Bellman operator is a span contraction \cite{MDPBook}. Based on the computed $(g^*, \bm{h}^*)$, the optimal policy $\mu^*$ can be extracted from $\bm{h}^*$ by acting greedy (i.e., choose the action that gives the minimal future cost), giving,
\begin{eqnarray*}
\mu^*(s) = \arg\min_{a}\left\{C(s)+\sum_{s'}P(s'|s,a) \bm{h}^*[s']  \right\},
\end{eqnarray*}

Relative value iteration gives us the optimal solution to \eqref{eq:II_opt_policy}, but it also presents several problems.
\begin{enumerate}
\item To compute \eqref{eq:III_bellman_operator}, the number of states of the MDP must be finite so that $\sum_{s'}P(s'|s,a)=1$. However, the state size in our problem is infinite, because the value of a counter can be any non-negative integers, i.e., $n_i\in\mathbb{N}^0$. As a result, we must set an upper limit, $U$, for each counter to enable the computation of \eqref{eq:III_bellman_operator} in each iteration (i.e., a counter value larger than $U$ is set to $U$). In order not to affect the optimality of the relative value iteration, $U$ must be set large enough so that $P(n_i > U)$ is negligible for a set of policies in the neighborhood of the optimal policy.
\item Given the upper limit $U$ for each counter, the state size is now $\left|S\right|=U^M$, and the decision space is $\left|S\right|\times \left|A\right|\times\left|S\right|=MU^{2M}$. The computational complexity of relative value iteration grows exponentially with the increase of $M$. This largely limits the scalability of the relative value iteration, and makes the optimal policy prohibitively expensive to compute for large $M$.
\item Relative value iteration solves the MDP by optimal planning. A prerequisite is that the controller must have perfect knowledge $p_i$, a parameter determined by the local volatility of each IoT device, to compute the transition probability $P(s'|s,a)$ in \eqref{eq:III_bellman_operator}. This prior information, however, may not be available to the controller in practice.
\end{enumerate}

In summary, a realistic and decent sampling policy should have 1) good performance in terms of minimizing the average cost, 2) low computational complexity, 3) very little reliance on the prior information of network dynamics. The above optimal policy does not satisfy the second and third requirements. In this context, we have to consider other low-complexity solutions that scale well with the number of IoT devices and do not rely on knowledge $p_i$.

\section{State-independent Policies}\label{sec:IV}
A sampling policy can be state-dependent or state-inde\-pendent. State-dependent policies, e.g., the optimal policy $\mu^*$ in \eqref{eq:II_opt_policy}, make the sampling decision based on the current states of the counters. State-independent policies, on the other hand, make the sampling decision regardless of the counter states. Compared with the optimal policy in \eqref{eq:II_opt_policy}, state-independent policies are suboptimal in general, but they have low complexity, hence are very easy to implement in practice. Moreover, some of the state-independent policies do not rely on prior information of the network parameters. This section focuses on two state-independent policies implemented in \cite{openTM}, i.e., the uniform policy and the non-uniform policy, and their generalizations. Their performance is analyzed in terms of the average cost over the infinite-time horizon.

A simple sampling strategy is uniform sampling. In each slot, a uniform policy samples one of the $M$ IoT devices on path $\overline{AB}$ uniformly at random. The performance of uniform sampling is characterized in Proposition \ref{thm:2}.

\begin{prop}[performance of the uniform policy]\label{thm:2}
The average cost of the uniform policy over the infinite-time horizon is given by
\begin{eqnarray}\label{eq:IV_uniform}
J_\textup{uniform} = \sum_{i=1}^{M} \frac{\varphi_i (M-1)(1-p_i)}{M-(M-1)(1-p_i)}.
\end{eqnarray}
\end{prop}

\begin{NewProof}
See Appendix \ref{sec:AppB}.
\end{NewProof}

Consider a homogeneous network wherein $p_1=p_2=...=p_M=p$ and let $\varphi_i=\sigma^{M-i}$. Eq.~\eqref{eq:IV_uniform} can be simplified to
\begin{eqnarray}\label{eq:IV_uniform2}
J_\textup{uniform} = \frac{(1-\sigma^M)(M-1)(1-p)}{(1-\sigma)[M-(M-1)(1-p)]}.
\end{eqnarray}
Eq. \eqref{eq:IV_uniform2} suggests that $J_\textup{uniform}$ monotonically increases with $M$. Let $M\rightarrow \infty$,
\begin{eqnarray*}
\lim_{M\rightarrow\infty} J_\textup{uniform} = \frac{1-p}{(1-\sigma)p}.
\end{eqnarray*}

The uniform policy samples each device with the same probability $1/M$ in each slot. Obviously, this policy is suboptimal when we have a sampling preference over different devices (i.e., when different IoT devices have different $\varphi_i$ and $p_i$). To tackle this problem, \cite{openTM} further proposed a non-uniform sampling policy: in each slot, the controller randomly generates two random integers between $1$ and $M$ with replacement, and then samples the device indexed by the larger integer. By so doing, the devices closer to the destination are more likely to be sampled. This matches the system model in \cite{openTM} because the controller is more inclined to sample the devices closer to the destination as they yield more accurate statistics. This non-uniform sampling policy can be generalized as follows.

\begin{defi}[largest-order-statistic policy]
In each slot, the largest-order-statistic policy randomly generates $G$ integers between $1$ and $M$ with replacement, and samples the device indexed by the largest integer (i.e., the largest order statistic of the uniform distribution).
\end{defi}

With the largest-order-statistic policy, the $M$ IoT devices are sampled by path $\overline{AB}$ in a non-uniform manner. Proposition \ref{thm:3} gives the performance of such a scheme.

\begin{prop}[performance of the largest-order-statistic policy]\label{thm:3}
The average cost of the largest-order-statistic policy over the infinite-time horizon is given by
\begin{eqnarray}\label{eq:IV_order}
J_\textup{order} = \sum_{i=1}^{M} \frac{\varphi_i (1-q_i)(1-p_i)}{1-(1-q_i)(1-p_i)},
\end{eqnarray}
where $q_i=\frac{i^G-(i-1)^G}{M^G}$.
Let $p_1=p_2=\cdots=p_M=p$ and $\varphi_i=\sigma^{M-i}$. If $M\rightarrow\infty$ and $M\gg G$, $J_\textup{order}$ converges to the same value as the uniform policy, giving
\begin{eqnarray}\label{eq:IV_order2}
\lim_{M\rightarrow\infty}J_\textup{order} = \frac{1-p}{(1-\sigma)p}.
\end{eqnarray}
\end{prop}

\begin{NewProof}
See Appendix \ref{sec:AppB}.
\end{NewProof}

The largest-order-statistic policy is better than the uniform policy in that the sampling distribution takes the different $\varphi_i$ and $p_i$ of different devices into account. A natural question is that, what is the optimal stationary state-independent policy? Said in another way, what is the optimal sampling distribution over the $M$ IoT devices that minimizes the average cost? This leads us to a weighted-probability sampling policy.

\begin{prop}[weighted-probability sampling policy]\label{thm:4}
In each time slot, a weighted-probability sampling policy samples the $i$-th device on the path $\overline{AB}$ with probability $w^*_i$, where
\begin{eqnarray*}
w^*_i = \left(v\sqrt{\frac{\varphi_i}{1-p_i}} -\frac{p_i}{1-p_i}  \right)^+,
\end{eqnarray*}
and $v$ is chosen such that
\begin{eqnarray*}
\sum_{i=1}^{M} \left(v\sqrt{\frac{\varphi_i}{1-p_i}} -\frac{p_i}{1-p_i}  \right)^+ = 1.
\end{eqnarray*}

The average cost over the infinite-time horizon achieved by the weighted-probability policy is
\begin{eqnarray}\label{eq:IV_weighted}
J_\textup{weighted} = \sum_{i=1}^M \varphi_i \left(\frac{1}{(1-p_i)w^*_i+p_i} - 1 \right).
\end{eqnarray}
\end{prop}

\begin{NewProof}
See Appendix \ref{sec:AppC}.
\end{NewProof}

Given the optimized sampling distribution to minimize the average cost, the weighted-probabi\-li\-ty sampling policy is the best strategy within the class of non-uniform sampling strategies. Notice that the average cost achieved by the weighted-probability policy is exactly twice the lower bound given in Theorem \ref{thm:1}.

\section{Index Policies}\label{sec:V}
This section considers a class of index policies to solve the MDP associated with the flow sampling problem. We will first leverage the Whittle index, originally proposed for restless multi-armed bandit (RMAB) problems, to devise a Whittle index sampling policy that has close-to-optimal performance and linear complexity in the problem size. Yet, as the optimal policy does, the Whittle index policy relies on the accurate estimation of $p_i$. Inspired by the form of the Whittle index, we put forth a second-order index policy at the end of this section. While inheriting all the advantages of the Whittle index, the second-order index does not require accurate prior-information of $p_i$.

\subsection{The decoupled problem}
Faced with an $M$-dimensional MDP, it is inevitable that the computational complexity of the optimal policy increases exponentially with the number of devices $M$. A possible scheme to admits linear-complexity policy is to decouple the $M$-dimensional problem to $M$ one-dimensional problems. Decoupling is the main idea of a series of index policies to solve the MAB problems.

When sampling the $M$ IoT devices on a flow path, the state evolution of each device is a controlled Markov process independent from other devices. Specifically, the evolution of $n_i$ is controlled by the ``sample'' action (i.e., $n_i$ goes to $0$ once being sampled, and $n_i+1$ otherwise), and is irrelevant to how $n_j$, $j\neq i$ evolves.

Let us consider a decoupled problem of sampling only one device. To simplify the notations, we remove the subscript $i$ for all the definitions in Section \ref{sec:II} since there is only one device. The state of the device is then $s=\{n: n=\mathbb{N}^0\}$, and the action space is $a=\{0,1\}$ where $0$ and $1$ correspond to ``rest'' and ``sample'', respectively. The state transition probability is given by
\begin{eqnarray*}
\left\{
\begin{array}{lll}
P\left(s^{t+1}=0 \left|\right. s^t=n, a^t=1 \right) = 1, & \\
P\left(s^{t+1}=0 \left|\right. s^t=n, a^t=0 \right) = p, & \\
P\left(s^{t+1}=n+1 \left|\right. s^t=n, a^t=0 \right) = 1-p, &
\end{array}
\right.
\end{eqnarray*}

The immediate cost incurred by being in state $s^t$ and executing $a^t$ is
\begin{eqnarray*}
\left\{
\begin{array}{lll}
C\left(s^t=n, a^t=1 \right) = c+\varphi n, & \\
C\left(s^t=n, a^t=0 \right) = \varphi n, &
\end{array}
\right.
\end{eqnarray*}
where $\varphi$ is the accuracy associated with this device, and $c\geq 0$ is a fixed sampling cost (defined later).

The optimal policy $\overline{\mu}^*$ for the decoupled problem is defined as
\begin{eqnarray}\label{eq:V_opt_policy}
\overline{\mu}^*=\arg\min_{\overline{\mu}} \lim_{T\rightarrow\infty} \mathbb{E} \left[\frac{1}{T}\sum_{t=0}^{T-1}C(s^t,a^t) \right].
\end{eqnarray}

Compared with the original $M$-device sampling problem, the decoupled problem introduces a fixed sampling cost $c$. Without this fixed sampling cost, the controller would keep sampling the device to minimize \eqref{eq:V_opt_policy}. To avoid this, we artificially introduce a fixed cost $c$ for each sampling operation. As per Whittle's argument, we aim to find the sampling cost $c^*$ for which it is equally optimal to ``sample'' and ``rest'' (i.e., the expected costs incurred by ``sample'' and ``rest'' are the same). In doing so, $c^*$, i.e., the Whittle index, acts as a measurement of how much the controller is willing to pay to sample this device.

In the original $M$-device sampling problem, we could compute the corresponding Whittle index for individual devices in each decision epoch, and sample the device with the largest Whittle index.

\subsection{Solving the Decoupled Problem}
The decoupled problem is also a controlled MDP. Given a sampling cost $c$, the optimal solution to the decoupled problem can be modified from \eqref{eq:III_bellman} as
\begin{eqnarray}\label{eq:V_bellman}
&&\hspace{-0.4cm} g^* + \bm{h}^*[n] =  \min \big\{c+\varphi n + \bm{h}^*[0], \nonumber\\
&&\hspace{1cm} \varphi n + p \bm{h}^*[0] + (1-p) \bm{h}^*[n+1] \big\},
\end{eqnarray}
where the two terms inside the minimization operation correspond to the costs incurred by the actions ``sample'' and ``rest'', respectively. Without loss of generality, we choose state $n=0$ as the reference state and set $\bm{h}^*[0]=0$. Thus,
\begin{eqnarray}\label{eq:V_bellman2}
\bm{h}^*[n]= \varphi n + \min \left\{c, (1-p) \bm{h}^*[n+1] \right\} - g^*.
\end{eqnarray}
Eq.~\eqref{eq:V_bellman2} defines the relative value function of each state $n$ under the optimal policy for a given sampling cost $c$.

\begin{prop}[solution to the decoupled problem]\label{thm:5}
The optimal policy $\overline{\mu}^*$ to the decoupled problem is a threshold policy. For a given sampling cost $c$, there exists an integer threshold $\Gamma(c)$ such that 1) if a state $n<\Gamma(c)$, the optimal policy is to ``rest'', and 2) if a state $n\geq\Gamma(c)$, the optimal policy is to ``sample''.
\end{prop}

\begin{NewProof}
See Appendix~\ref{sec:AppE}.
\end{NewProof}

\begin{figure}[t]
  \centering
  \includegraphics[width=0.9\columnwidth]{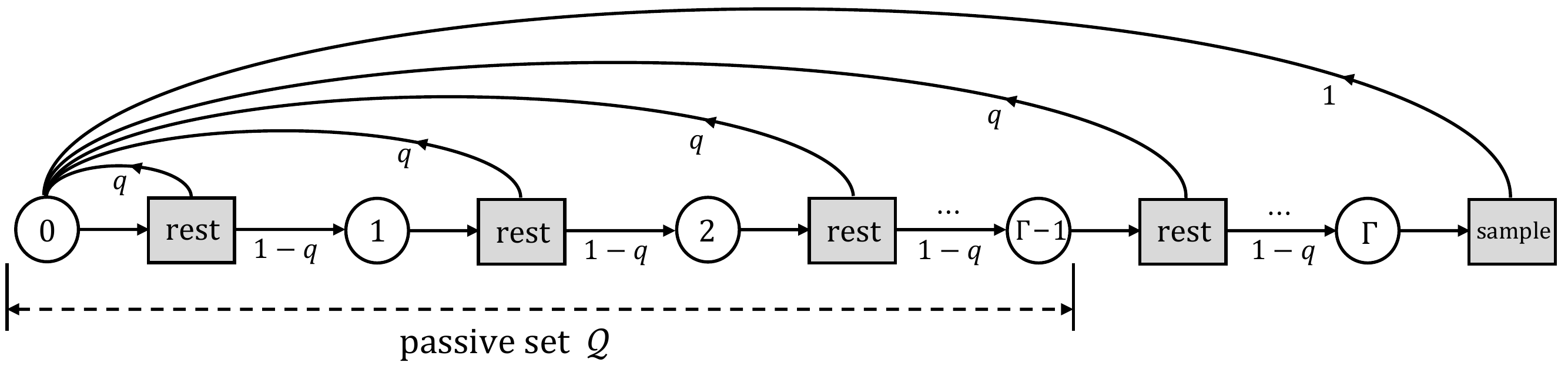}\\
  \caption{Under the threshold policy, the decoupled problem is a unichain with a single recurrent class. All the states $n>\Gamma$ are transient states. The circles in the figure are states, while the rectangles are actions.}
\label{fig:3}
\end{figure}

Given the threshold structure of the optimal policy, the decoupled problem is essentially a unichain with a single recurrent class. In equilibrium, the transitions of recurrent states are illustrated in Fig.~\ref{fig:3}. We define the set of states wherein the optimal policy is ``rest'' as the ``passive set'', i.e.,
\begin{eqnarray}\label{eq:V_passiveset}
\mathcal{Q}(c) = \{n: 0\leq n<\Gamma(c), n\in\mathbb{Z}   \}
\end{eqnarray}

\subsection{Whittle index policy}
Whittle index is a good heuristic to solve RMAB problems provided that the problem is indexable. As noted by Whittle \cite{Whittle1988}, a decoupled problem is said to be indexable if the passive set $\mathcal{Q}(c)$ is monotone non-decreasing as the subsidy (in our case, sampling cost) increases. That is, for any real values $c_1<c_2$, the passive set $\mathcal{Q}(c_1)\subseteq\mathcal{Q}(c_2)$. An RMAB problem is indexable if all its arms are indexable.

\begin{lem}[Indexability]\label{thm:6}
The decoupled problem in \eqref{eq:V_opt_policy} as well as the original M-device sampling problem in \eqref{eq:II_opt_policy} are indexable.
\end{lem}

\begin{NewProof}
Let $n=\Gamma-1$ and $n=\Gamma$ in \eqref{eq:V_F1} and \eqref{eq:V_F4}, respectively, we have
\begin{eqnarray}\label{eq:V_F12}
\bm{h}^*[\Gamma] \leq \frac{c}{1-p} \leq \bm{h}^*[\Gamma+1].
\end{eqnarray}

Given a sampling cost $c$, Eq. \eqref{eq:V_F12} means there exists one and only one $\Gamma(c)$ such that $\frac{c}{1-p}$ falls into the interval $\left[ \bm{h}^*[\Gamma(c)], \bm{h}^*[\Gamma(c)+1] \right]$.

From the proof of Proposition \ref{thm:5}, we know that $\bm{h}^*[n]$ is a strictly increasing function of $n$. Thus, $\Gamma(c)$ is monotone nondecreasing in $c$ (it is a staircase function since $\Gamma$ takes integer values), and the passive set $\mathcal{Q}(c)$ defined in \eqref{eq:V_passiveset} is monotone nondecreasing in $c$.

As a result, the decoupled problem for each device is indexable, hence the original M-device sampling problem in \eqref{eq:II_opt_policy} is also indexable.
\end{NewProof}

Given the indexability condition established in Lemma \eqref{thm:6}, the Whittle index policy is captured by Theorem \eqref{thm:7} below:

\begin{thm}[Whittle index policy]\label{thm:7}
At the beginning of a slot $t$, the controller computes a Whittle index $c^*(n_i)$ separately for each device as a function of its current state $n_i$, and then samples the device with the greatest index. The whittle index is given by
\begin{eqnarray}\label{eq:V_F13}
c^*(n_i) = \frac{\varphi_i(1-p_i)}{p^2_i}\left[ (1-p_i)^{n_i+2} + (n_i+2)p_i - 1 \right].
\end{eqnarray}
\end{thm}

\begin{NewProof}
See Appendix \ref{sec:AppF}.
\end{NewProof}

When in state $n_i$, the Whittle index $c^*(n_i)$ measures how attractive is the $i$-th device to the controller.

\begin{figure}[t]
  \centering
  \includegraphics[width=0.8\columnwidth]{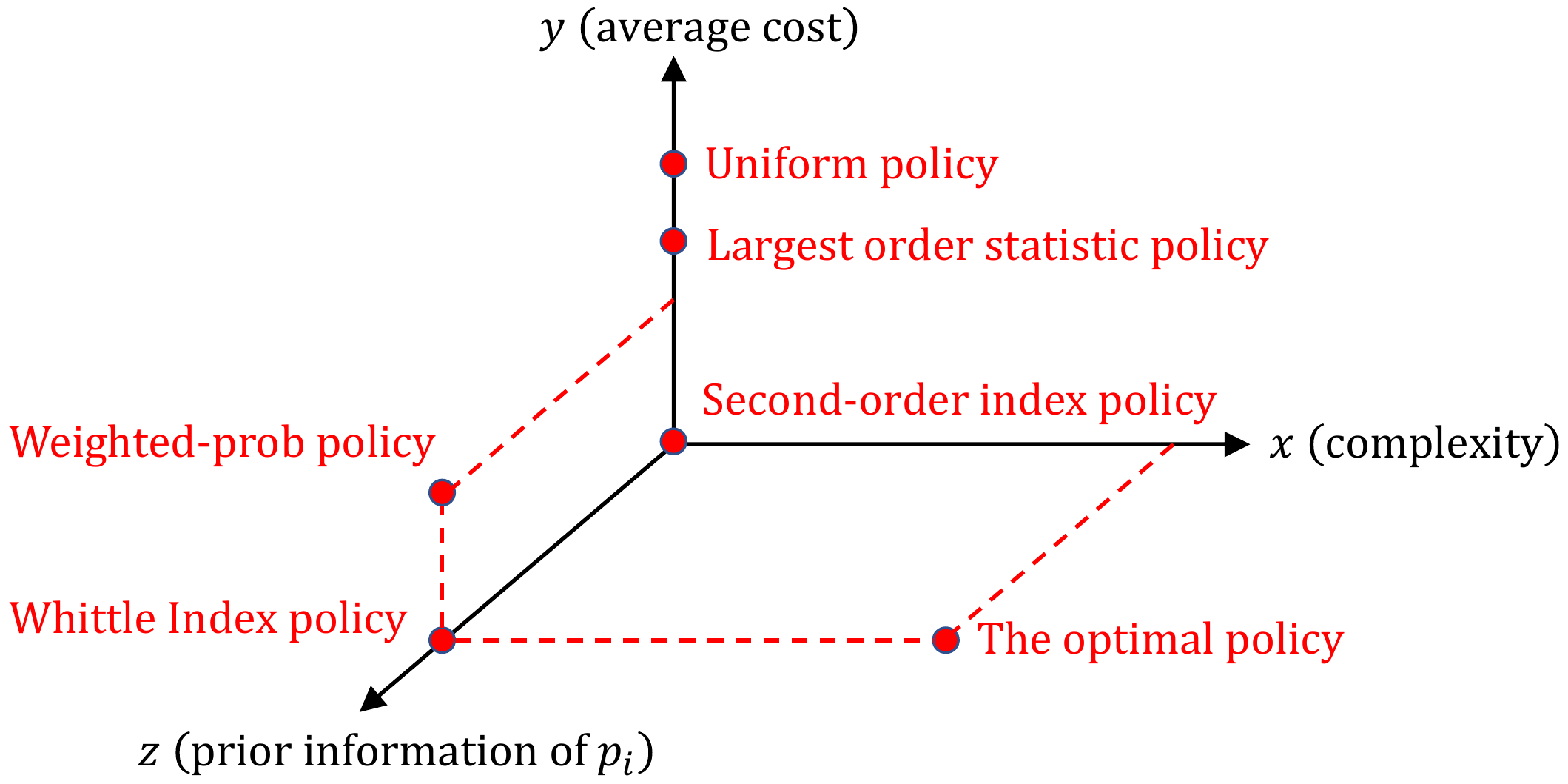}\\
  \caption{Comparisons among the optimal policy, the state-independent policies (uniform, largest-order-statistic, and weighted-probability policies), the Whittle index policy, and the second-order index policies in terms of the computation complexity ($x$-axis), the performance of average cost ($y$-axis), and the requirements of prior knowledge $p_i$ (z-axis). The average costs of different policies can be found in Section \ref{sec:VI}. The relative performance of different policies is provided for illustrative purposes only and is not meant to depict the precise performance gaps.}
\label{fig:4}
\end{figure}

\subsection{The second-order index policy}
In Fig. \ref{fig:4}, the optimal policy, the state-independent policies, and the Whittle index policy are evaluated in a three-dimensional coordinate system. The positive direction of the $x$-axis means the policy requires higher computational complexity, the positive direction of the $y$-axis means the policy yields larger average cost (poorer performance), and the positive direction of the $z$-axis means the policy requires prior-information $p_i$, a parameter determined by the local volatility of each device.

As shown, the Whittle index policy is preferred to the optimal policy and the state-independent policies thanks to its low complexity and decent average-cost performance. Yet, the execution of the Whittle index policy hinges on the accurate estimation of $p_i$, as the optimal policy does. Since the accurate estimates of $p_i$ may not be available to the controller in practice, we put forth a second-order index policy in the following that does not rely on prior information $p_i$, while inheriting all the advantages of the Whittle index.

\begin{defi}[second-order index policy]
At the beginning of a slot $t$, the controller computes a second-order index $I(n_i)$ separately for each device as a function of its current state $n_i$, and then samples the device with the greatest index. For the $i$-th device, the second-order index is given by
\begin{eqnarray}\label{eq:V_2order}
I(n_i) =\lim_{p_i\rightarrow 0} c^*(n_i) = \frac{\varphi_i}{2} (n_i+1) (n_i+2).
\end{eqnarray}
\end{defi}

It is plausible that the second-order index policy performs well when $p_i$,$\forall i$ are small, because the second-order index is inferred from the Whittle index by assuming a device undergoes very light traffic with $p_i\rightarrow 0$. However, one may ask, does this second-order index perform well when some of the IoT devices undergo moderate or heavy traffic with relatively large $p_i$? We answer this question affirmatively by the simulation results in section \ref{sec:VI}, where it is shown that the second-order index policy performs well for both small and large $p_i$.

Overall, the second-order index policy is the most desired policy among other policies. As shown in Fig.~\ref{fig:4}, it has low computation complexity, no reliance on the prior-information $p_i$, and comparable average-cost performance to the Whittle index policy.

\begin{rem}
When $p_i$ of the $i$-th device is large, an alternative to the second-order index in \eqref{eq:V_2order} is a first-order index
\begin{eqnarray}\label{eq:V_1order}
I(n_i) =\lim_{p_i\rightarrow 1} \frac{c^*(n_i)}{1-p_i} = \varphi_i (n_i+1).
\end{eqnarray}
This gives us the following heuristic index policy.

\textbf{Heuristic index policy} -- Assume the controller has a rough idea of whether $p_i$ is larger or smaller than a threshold probability $\overline{p}$ for each device. At a decision epoch, the controller takes the second-order index in \eqref{eq:V_2order} as the heuristic index for IoT devices whose $p_i<\overline{p}$; and the first-order index in \eqref{eq:V_1order} as the heuristic index for IoT devices whose $p_i\geq \overline{p}$. Then, the controller samples the device with the largest heuristic index.

This heuristic index policy is evaluated at the end of section \ref{sec:VI}. It is shown that the heuristic index policy only yields minor gains over the second-order policy. Yet, it requires the controller to know a certain amount of prior knowledge $p_i$, and the threshold probability $\overline{p}$ must be chosen very carefully. Overall, the second-order index is good enough to ensure a minor gap to the Whittle index policy.
\end{rem}

\section{Numerical and Simulation Results}\label{sec:VI}
\subsection{The optimal policy and the lower bound}
As stated in Section \ref{sec:III}, the computational complexity of relative value iteration is prohibitively high. This makes the optimal policy in \eqref{eq:II_opt_policy} very expensive to obtain, especially when the number of IoT devices $M$ is large. In view of this, we first consider a simple case where there are only three devices to evaluate the performance gap between the optimal policy and the lower bound given in Theorem \ref{thm:1}.

\begin{figure}[t]
  \centering
  \includegraphics[width=0.8\columnwidth]{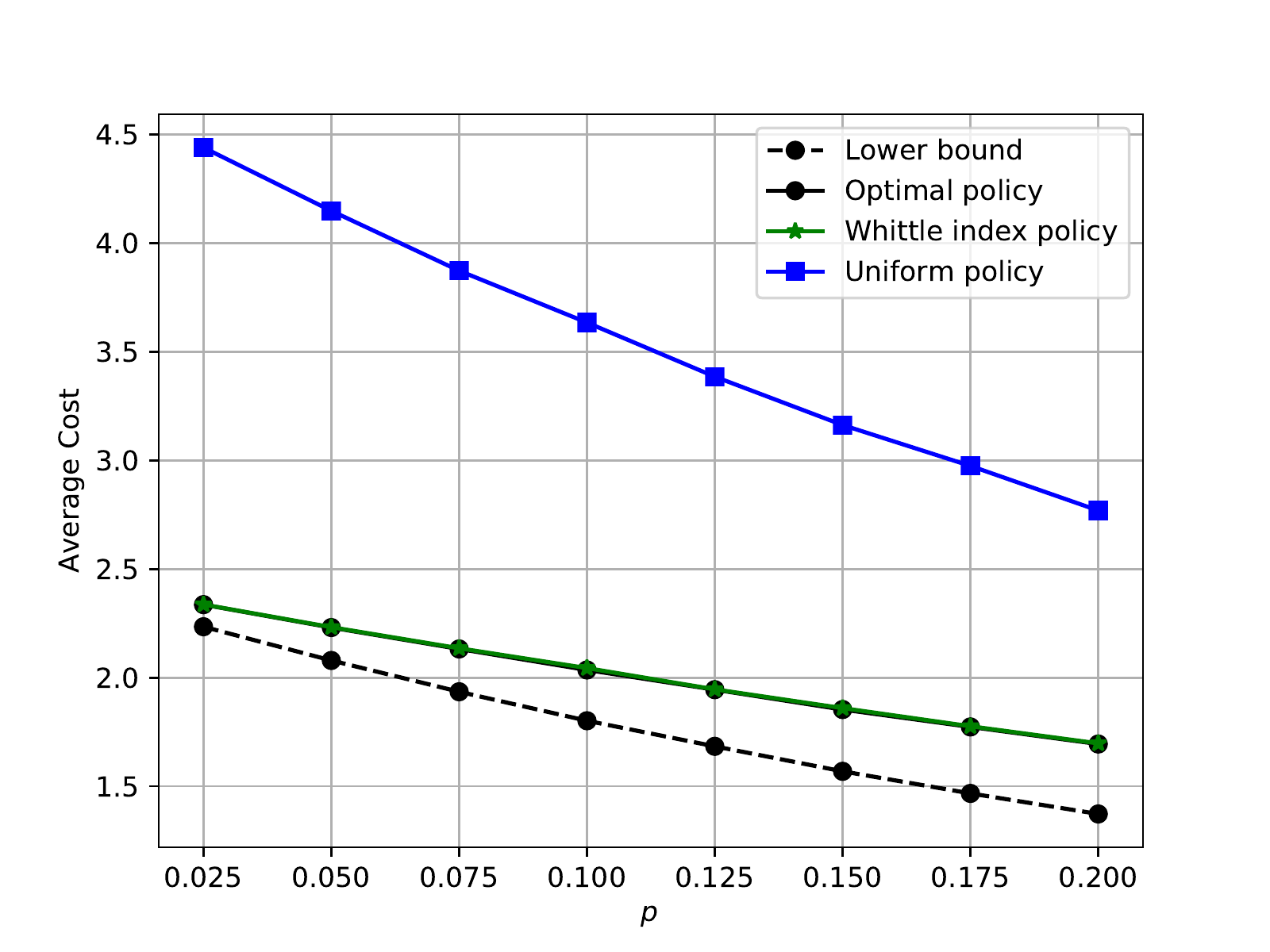}\\
  \caption{The performance of the optimal policy benchmarked against the lower bound, wherein $M = 3$. The performance of the uniform policy and the Whittle index policy is also plotted in this same figure.}
\label{fig:S1}
\end{figure}

Fig. \ref{fig:S1} presents the average costs achieved by the optimal policy, the uniform policy, and the Whittle index policy benchmarked against the lower bound on a flow path with $M = 3$ IoT devices. In this figure, we fix $\sigma=0.8$, i.e., the accuracies of statistics collected from the three IoT devices are $0.64$, $0.8$, and $1$, respectively. The probability that a device is sampled by flows other than $\overline{AB}$ is set to $p_1=p_2=p_3=p$, and we increase $p$ from $0.025$ to $0.2$. To execute relative value iteration and compute the optimal policy, we set the upper limit $U$ of each counter to $10$ (i.e., a counter no longer grows when it reaches $10$). The size of the state space is then $\left|\mathcal{S}\right|=U^M=1000$, and the decision space is $\left|\mathcal{S}\right|\times\left|\mathcal{A}\right|\times\left|\mathcal{S}\right|=\allowbreak MU^{2M}=\allowbreak 3\times 10^6$.

As can be seen from Fig.~\ref{fig:S1},
\begin{enumerate}
\item The Whittle index policy performs as well as the optimal policy for small $M$ (the two curves coincident with each other). However, the optimality of the Whittle index is unknown in the case of large $M$ due to the unavailability of the optimal policy.
\item The performance gap between the optimal policy and the lower bound is minor when $p$ is small, but gets larger as $p$ increases. This is not surprising because to derive the lower bound, we have assumed in Theorem \ref{thm:1} that the variance of the inter-sampling time of each device is negligible relative to the mean of the inter-sampling time. Thus, the lower bound is supposed to be tighter in the case of larger $M$ and smaller $p$.
\end{enumerate}

Assuming a large number of IoT devices, the following parts evaluate the performance of state-independent policies proposed in \cite{openTM} and our second-order index policy. Keeping in mind that the Whittle index policy can be suboptimal, and the lower bound may not be tight, we will take them as the benchmarks.

\subsection{State-independent policies}
This subsection evaluates the average costs achieved by different state-independent policies and their generalizations, i.e., the uniform sampling policy, the largest-order-statistic policy, and the weighted-probability policy.

\begin{figure}[t]
  \centering
  \includegraphics[width=0.9\columnwidth]{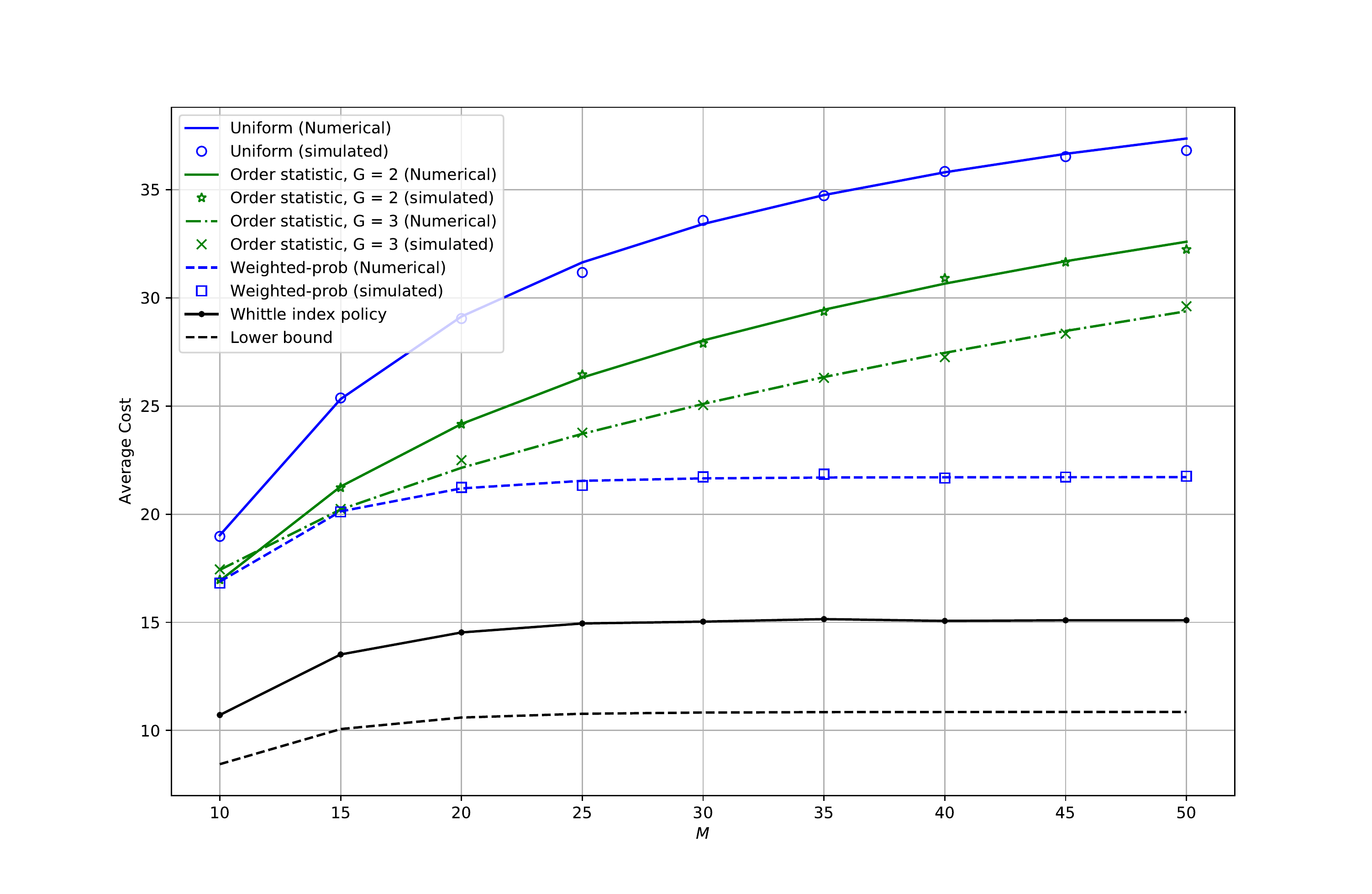}\\
  \caption{Numerical and simulation results for the uniform policy, the largest-order-statistic policy (with $G = 2$ and $3$), and the weighted-probability policy, wherein $\sigma=0.8$, $p_1=p_2=\cdots=p_M=p=0.1$. The lower bound and the performance of the Whittle index policy are plotted in the same figure.}
\label{fig:S2}
\end{figure}

The numerical and simulation results of the above three state-independent sampling policies are presented in Fig.~\ref{fig:S2}, where we fix $\sigma=0.8$, and $p_1=p_2=\cdots=p_M=p=0.1$. The analytical results match with the simulation results very well.

As can be seen from Fig.~\ref{fig:S2},
\begin{enumerate}
\item Uniform sampling gives the worst performance. The average cost, as predicted in \eqref{eq:IV_uniform2}, increases monotonically with the increase of $M$. As $M$ goes to infinity, the average cost converges to $\frac{1-p}{(1-\sigma)p}=45$.
\item The performance of the largest-order-statistic policy depends on the value of $G$, i.e., the number of random integers generated each time. For a fixed $G\ll M$, \eqref{eq:IV_order2} indicates that the average cost converges to the same value $\frac{1-p}{(1-\sigma)p}=45$ as the uniform policy.
\item The weighted-probability policy outperforms both the uniform policy and the largest-order-statistic policy. This outcome is expected because we have optimized the sampling probability over all IoT devices to devise the weighted-probability policy. As indicated in \eqref{eq:IV_weighted}, the performance of the weighted-probability policy is twice of the lower bound. With the increase of $M$, the average cost converges to around $22.64$.
\item The Whittle index policy outperforms all three state-independent policies. Compared with the uniform policy and the largest-order-statistic policy, the Whittle index policy reduces the average cost by $66.4\%$ when $M$ goes to infinity. Compared with the weighted-probability policy, the Whittle index policy reduces the average cost by $33.4\%$ when $M$ goes to infinity.
\end{enumerate}

\subsection{The Second-order index policy}
The Whittle index policy outperforms the state-independent policies by much, but it requires accurate estimates of $p_i$ to compute the indexes. An alternative to the Whittle index is the second-order index given in \eqref{eq:V_2order}, the computation of which does not require any prior information $p_i$. This subsection verified the performance of the second-order index policy benchmarked against the Whittle index policy.

We consider an asymmetric network where IoT devices undergo two kinds of sampling-request traffic: 1) all the odd-indexed devices undergo light traffic with small $p_i=\pi_0$; and 2) all the even-indexed devices undergo moderate/heavy traffic with relatively large $p_i=\pi_1$. In the simulation, we fix $\pi_0$ to $0.01$, and vary $\pi_1$. For the Whittle index policy, $\pi_0$ and $\pi_1$ are assumed to be known to the controller such that the Whittle index can be computed. For the second-order index policy, the controller computes the second-order index directly from \eqref{eq:V_2order}.

\begin{figure}[t]
  \centering
  \includegraphics[width=0.8\columnwidth]{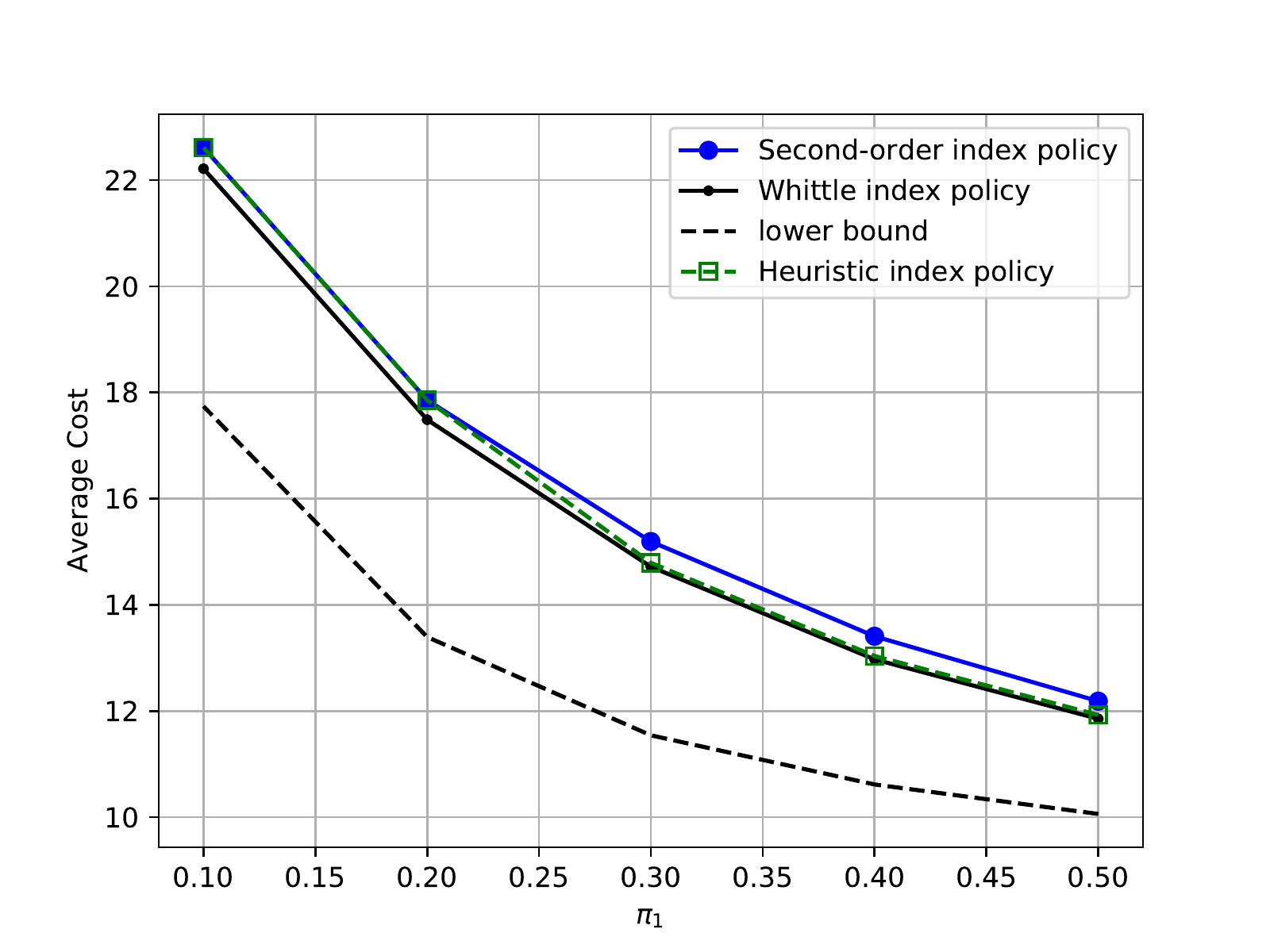}\\
  \caption{Performance comparison between the second-order index policy and the Whittle index policy, wherein $M = 40$ and $\sigma=0.8$.}
\label{fig:S3}
\end{figure}

Fig.~\ref{fig:S3} presents the average costs achieved by the second-order index and the Whittle index policies in the considered network, wherein $M = 40$. As shown, for different $\pi_1$, the performance gaps between the two policies are very small. The second-order index policy is a good substitute for the Whittle index policy given the same low-complexity property and comparable average-cost performance. Better yet, the second-order index policy requires no prior-information of $p_i$.

Finally, we evaluate the heuristic index policy in the same network. As per the heuristic index policy, the controller has to compute a heuristic index for each device in a decision epoch. To this end, we first set the threshold probability $\overline{p}=0.3$. That is, the heuristic index of the $i$-th device is the second-order index given in \eqref{eq:V_2order} if $p_i<0.3$, and is the first-order index given in \eqref{eq:V_1order} if $p_i\geq 0.3$. The controller then samples the device with the largest heuristic index.

The performance of the heuristic index policy is plotted in Fig.~\ref{fig:S3}. As shown, when $\pi_1<0.3$, the performance of the heuristic index policy is the same as that of the second-order index, because all $p_i$ in the network are smaller than the threshold probability $0.3$. On the other hand, when $\pi_1\geq 0.3$, the indexes of all even-indexed IoT devices are the first-order indexes rather than the second-order indexes. The heuristic index policy is slightly better than the second-order index policy. However, the downsides are that the controller has to know a certain amount of information about $p_1$, and the threshold probability $\overline{p}$ must be chosen very carefully (an ill-chosen $\overline{p}$ easily leads to substantial performance degradations).

\section{Conclusion}\label{sec:Conclusion}
In software-defined Internet-of-Things networking (SDIoT), the controller samples each active flow to gather network information for traffic engineering and management.
A good sampling policy should sample the IoT devices to meet the controller's sampling preference and balance the query loads on the IoT devices.
In addition, a practical sampling policy should be computation-friendly, and has little reliance on prior knowledge of the network dynamics since they may be unavailable in practice.
The policies that meet these requirements, to our knowledge, are lacking in the literature.

To fill this research gap, this paper investigated the flow sampling problem in large-scale SDIoT networks, and studied the performance of different policies with the above criteria. Our main contributions are as follows:
\begin{enumerate}
\item We formulated the flow sampling problem in SDIoT networks by a Markov decision process (MDP). The optimal policy to this MDP is defined as the policy that makes the best tradeoffs between sampling accuracy and load balancing among IoT devices. We solved the MDP by a relative value iteration algorithm and derived the optimal policy.

\item We analyzed two state-independent policies previously proposed by others and generalized them to a largest-order-statistic policy and a weighted-probability policy. The weighted-probabi\-li\-ty policy was shown to be the optimal stationary state-independent policy. The performance of these policies was derived and validated by simulation results.

\item We transformed the MDP into a restless multi-armed bandit (RMAB) problem that admits a Whittle index policy. The closed-form Whittle index was derived. The Whittle index policy is near-optimal and has better performance than the previously proposed state-independent policies and their generalizations. The Whittle index policy, however, requires prior knowledge of the network dynamics.

\item Inspired by the Whittle index policy, we put forth a second-order index policy. This policy meets all the expectations we have for a practical policy: it is easy to compute, strikes very good tradeoffs between sampling accuracy and load balancing, and does not require any prior knowledge of the network dynamics.
\end{enumerate}

\appendices
\section{A Lower Bound to The Average Cost}\label{sec:AppA}
This appendix proves Theorem \ref{thm:1}.

Let us focus on the $i$-th counter $n^t_i$, and study how it evolves. As per \eqref{eq:II_tran1} and \eqref{eq:II_tran2}, $n^t_i$ is reset to $0$ once the $i$-th device is sampled. In between two sampling slots of the $i$-th device, $n^t_i$ increases from $0$ to $d-1$ if the inter-sampling time is $d$ slots.

Following the sample-path analysis \cite{samplePath,Kadota1}, we consider one sampling trajectory of the controller, and assume the $i$-th device is sampled at slot $t_{i,1}$, $t_{i,2}$, $\cdots$, $t_{i,K_i}$. As $K_1,\allowbreak K_2,\allowbreak \cdots,\allowbreak K_M\rightarrow\allowbreak \infty$, $t_{1,K_1}=\allowbreak t_{2,K_2}=\allowbreak\cdots=\allowbreak t_{M,K_M}=T\rightarrow\allowbreak \infty$. The sample mean and sample variance of the inter-sampling time $d_{i,k}=t_{i,k}-t_{i,k-1}$ ($k=1,2,...,K_i$, $t_{i,0}=0$) are defined as
\begin{eqnarray*}
&&\hspace{-0.2cm}\mathbb{E}[d_i] = \frac{1}{K_i}\sum_{k=1}^{K_i} d_{i,k}, \\
&&\hspace{-0.2cm}\mathbb{V}[d_i] = \frac{1}{K_i}\sum_{k=1}^{K_i} \left(d_{i,k} - \mathbb{E}[d_i] \right)^2= \frac{1}{K_i}\sum_{k=1}^{K_i} d^2_{i,k} - \mathbb{E}^2[d_i]. \\
\end{eqnarray*}

The average cost $J_\mu$ in \eqref{eq:II_avecost} can then be manipulated as follows:
\begin{eqnarray}\label{eq:A_1}
\hspace{-0.2cm}J_\mu \hspace{-0.3cm}&&\hspace{-0.2cm}= \lim_{T\rightarrow\infty} \mathbb{E}_\mu \left[\sum_{i=1}^{M}\varphi_i\frac{1}{T}\sum_{t=0}^{T-1}n^t_i \right] \\
&&\hspace{-0.2cm}= \lim_{T,K_i\rightarrow\infty} \mathbb{E}_\mu  \left[\frac{1}{T}\sum_{i=1}^{M}\varphi_i\sum_{k=1}^{K_i}\sum_{n=0}^{d_{i,k}-1}n \right]       \nonumber\\
&&\hspace{-0.2cm} = \lim_{T,K_i\rightarrow\infty} \mathbb{E}_\mu  \left[\frac{1}{2T}\sum_{i=1}^{M}\varphi_i\sum_{k=1}^{K_i}(d^2_{i,k}-d_{i,k}) \right] \nonumber\\
&&\hspace{-0.2cm} = \lim_{T,K_i\rightarrow\infty} \frac{1}{2T}\sum_{i=1}^{M}\varphi_i K_i \left(\mathbb{V}[d_i]+\mathbb{E}^2[d_i]-\mathbb{E}[d_i]\right).\nonumber
\end{eqnarray}

Since $\mathbb{E}[d_i]=T/K_i$ as $T,K_i\rightarrow\infty$, we have
\begin{eqnarray}\label{eq:A_2}
\hspace{-0.2cm}J_\mu \hspace{-0.3cm}&&\hspace{-0.2cm}= \lim_{T,K_i\rightarrow\infty} \frac{1}{2T}\sum_{i=1}^{M}\varphi_i K_i \left(\mathbb{V}[d_i]+\frac{T^2}{K^2_i}-\frac{T}{K_i}\right) \nonumber\\
&&\hspace{-0.2cm} = \lim_{T,K_i\rightarrow\infty} \frac{1}{2}\sum_{i=1}^{M}\varphi_i \left(\frac{K_i}{T}\mathbb{V}[d_i]+\frac{T}{K_i}-1\right) \nonumber\\
&&\hspace{-0.2cm}\geq \lim_{T,K_i\rightarrow\infty} \frac{1}{2}\sum_{i=1}^{M}\varphi_i \left(\frac{T}{K_i}-1\right).
\end{eqnarray}

Eq.~\eqref{eq:A_2} gives us a lower bound on $J_\mu$:
\begin{eqnarray}\label{eq:A_3}
L_B(J_\mu) = \min \lim_{T,K_i\rightarrow\infty} \frac{1}{2}\sum_{i=1}^{M}\varphi_i \left(\frac{T}{K_i}-1\right).
\end{eqnarray}

This bound can be further refined as follows.

\begin{figure}[t]
  \centering
  \includegraphics[width=0.8\columnwidth]{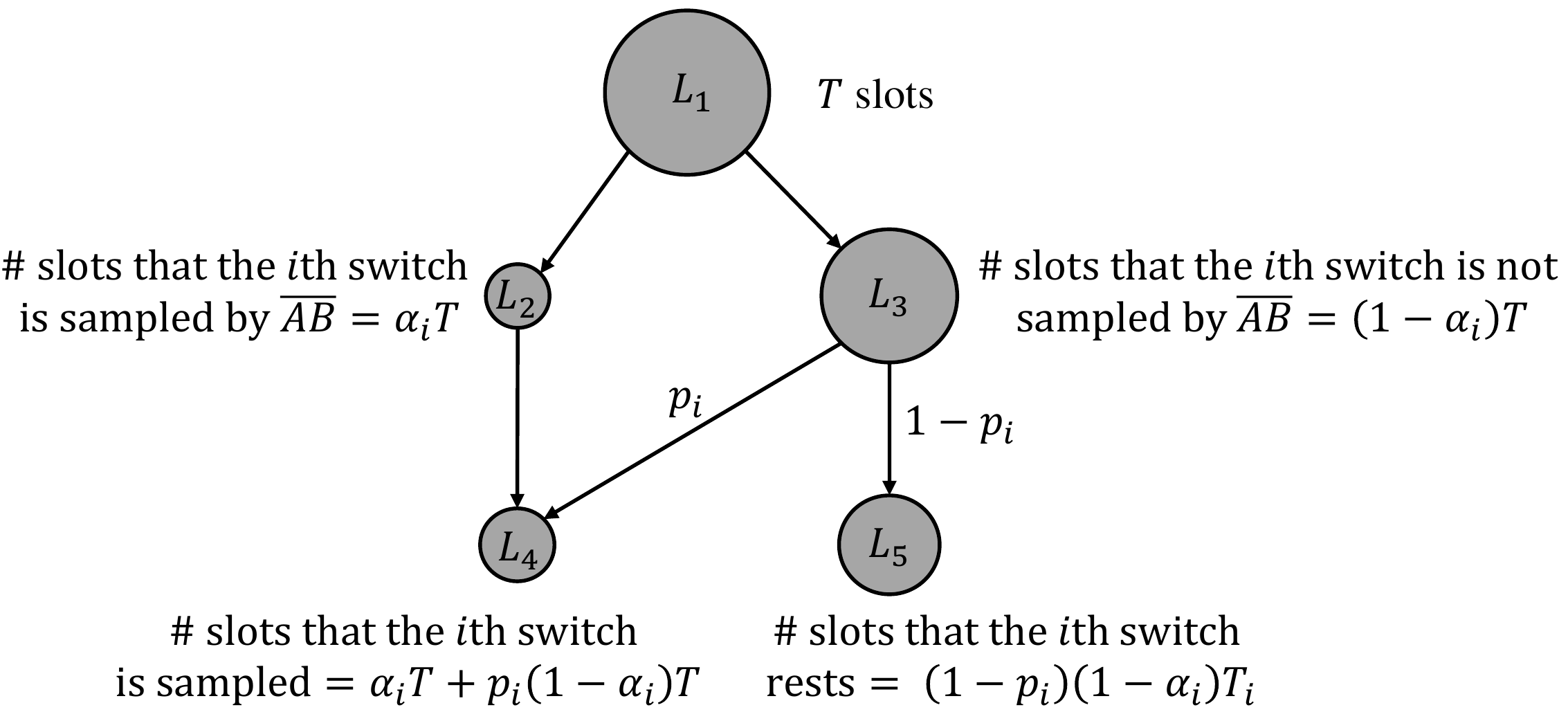}\\
  \caption{The behavior of the $i$-th device in one sampling trajectory.}
\label{fig:P1}
\end{figure}

The behavior of the $i$-th device in one sampling trajectory can be understood from Fig.~\ref{fig:P1}. As shown, there are overall $T$ slots, among which the i-th device is sampled by flow $\overline{AB}$ for $\alpha_i T$ slots for some $\alpha_i>0$. For the other $(1-\alpha_i)T$ slots, the $i$-th device can be sampled by flow paths other than $\overline{AB}$, or simply rest. Overall, the number of slots that the $i$-th device is sampled is given by
\begin{eqnarray}\label{eq:A_4}
K_i = \alpha_i T + p_i (1-\alpha_i) T.
\end{eqnarray}

Also, we have
\begin{eqnarray}\label{eq:A_5}
\sum_{i=1}^M \alpha_i T = T,
\end{eqnarray}
because flow path $\overline{AB}$ samples a device in every slot.

Substituting \eqref{eq:A_4} into \eqref{eq:A_3}, we have
\begin{eqnarray}\label{eq:A_6}
L_B(J_\mu) = \min \frac{1}{2}\sum_{i=1}^{M}\varphi_i\left(\frac{1}{(1-p_i)\alpha_i+p_i} - 1\right).
\end{eqnarray}

Eq. \eqref{eq:A_5} and \eqref{eq:A_6} give us the following linear program:
\begin{eqnarray}\label{eq:A_7}
&&\hspace{-0.2cm} L_B(J_\mu) = \min \frac{1}{2}\sum_{i=1}^{M}\varphi_i\left(\frac{1}{(1-p_i)\alpha_i+p_i} - 1\right) \nonumber\\
&&\hspace{-0.2cm} s.t.  \sum_{i=1}^M \alpha_i = 1,~~\alpha_i \geq 0.
\end{eqnarray}

Form the Lagrangian as
\begin{eqnarray}\label{eq:A_8}
&&\hspace{-0.2cm} f(\alpha_i, \lambda_0,\lambda_i) =  \frac{1}{2}\sum_{i=1}^{M}\varphi_i\left(\frac{1}{(1-p_i)\alpha_i+p_i} - 1\right) \nonumber\\
&&\hspace{2cm} + \lambda_0\left(\sum_{i=1}^M \alpha_i-1\right) - \lambda_i \alpha_i.
\end{eqnarray}

The optimal solutions $\{\alpha^*_i: i=1,2,\cdots,M \}$ to \eqref{eq:A_7} must satisfy the Karush-Kuhn-Tucker (KKT) conditions \cite{KKT} as follows.
\begin{eqnarray*}
&&\hspace{-0.2cm} \left.\frac{\partial f(\alpha_i, \lambda_0,\lambda_i)}{\partial \alpha_i} \right|_{\alpha^*_i} = \frac{-\varphi_i(1-p_i)}{2[(1-p_i)\alpha^*_i+p_i]^2} +  \lambda_0 - \lambda_i = 0,  \\
&&\hspace{-0.2cm} \left.\frac{\partial f(\alpha_i, \lambda_0,\lambda_i)}{\partial \lambda_0} \right|_{\alpha^*_i} =
\sum_{i=1}^M \alpha^*_i-1 = 0, \\
&&\hspace{-0.2cm}\lambda_i\alpha^*_i = 0, ~~~~\lambda_0 \geq 0, ~~~~\lambda_i \geq 0.
\end{eqnarray*}

A valid solution to \eqref{eq:A_7} is given by
\begin{eqnarray}\label{eq:A_12}
\alpha^*_i = \left(v\sqrt{\frac{\varphi_i}{1-p_i}} -\frac{p_i}{1-p_i}  \right)^+,
\end{eqnarray}
where $(x)^+=x$ if $x\geq 0$, and $(x)^+=0$ if $x< 0$; $v$ is chosen such that
\begin{eqnarray*}
\sum_{i=1}^{M} \left(v\sqrt{\frac{\varphi_i}{1-p_i}} -\frac{p_i}{1-p_i}  \right)^+ = 1.
\end{eqnarray*}

It is easy to verify that \eqref{eq:A_12} satisfies the KKT condition by setting
\begin{eqnarray*}
\lambda_0 = \frac{1}{2v^2} \geq 0,~~~~
\lambda_i = \frac{1}{2}\left(\frac{1}{v^2}-\frac{\varphi_i(1-p_i)}{p^2_i} \right) \geq 0.
\end{eqnarray*}
The lower bound is thus given by
\begin{eqnarray*}
L_B(J_\mu)=\frac{1}{2}\sum_{i=1}^M \varphi_i \left(\frac{1}{(1-p_i)\alpha^*_i+p_i} - 1 \right).
\end{eqnarray*}

\section{}\label{sec:AppB}
Under the uniform sampling policy, we can rewrite \eqref{eq:II_avecost} as
\begin{eqnarray}\label{eq:B_1}
\hspace{-0.5cm}&&J_\textup{uniform} = \lim_{T\rightarrow\infty} \mathbb{E}\left[\frac{1}{T}\sum_{t=0}^{T-1}\sum_{i=1}^{M}\varphi_i n^t_i \right] \nonumber\\
\hspace{-0.5cm}&&= \lim_{T\rightarrow\infty} \mathbb{E}\left[\sum_{i=1}^{M}\varphi_i\frac{1}{T}\sum_{t=0}^{T-1} n^t_i \right] = \sum_{i=1}^{M} \varphi_i \mathbb{E}[n_i],
\end{eqnarray}
where the last equality holds because the uniform policy is stationary, i.e., the probability that an IoT device is sampled does not change over time. Given \eqref{eq:B_1}, our target is to derive the average value of each counter.

For the $i$-th device, the probability that it is sampled by flow paths other than $\overline{AB}$ is $p_i$, and the probability that it is sampled by path $\overline{AB}$, denoted by $q_i$, is $1/M$. Thus, in any slot, the probability that the $i$-th device is not sampled by the controller is
$a_i = (1-q_i)(1-p_i)$
and $1-a_i$ is the probability that the $i$-th device is sampled in a slot.

\begin{figure}[t]
  \centering
  \includegraphics[width=0.8\columnwidth]{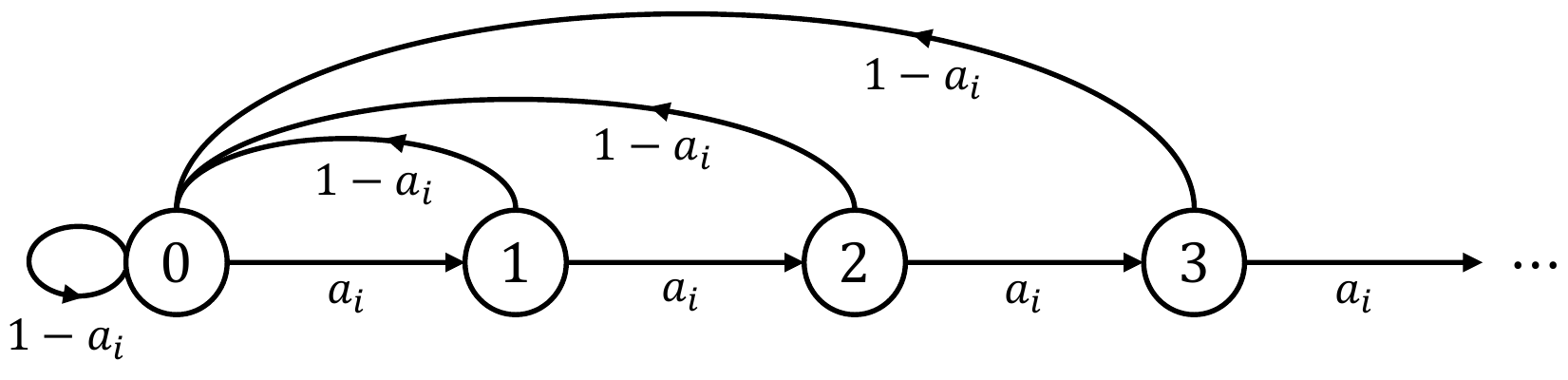}\\
  \caption{Under the uniform sampling policy, the Markov chain associated with the state transitions of one IoT device~\cite{IFDMA}.}
\label{fig:P2}
\end{figure}

As shown in Fig.~\ref{fig:P2}, the state transitions of the $i$-th device form a Markov Chain, the equilibrium distribution of which is given by
$\pi_n = a^n_i (1-a_i), n = 0, 1, 2, ...$
Thus,
\begin{eqnarray}
\label{eq:B_3}
&& \hspace{-1cm} \mathbb{E}[n_i] = \sum_{n=0}^{\infty} n \pi_n = \frac{a_i}{1-a_i}, \\
\label{eq:B_4}
&& \hspace{-1cm} J_\textup{uniform} = \sum_{i=1}^{M} \varphi_i \mathbb{E}[n_i]  \nonumber\\
&& \hspace{-1cm}  = \sum_{i=1}^{M} \varphi_i \frac{a_i}{1-a_i} = \sum_{i=1}^{M} \frac{\varphi_i(M-1)(1-p_i)}{M-(M-1)(1-p_i)}.
\end{eqnarray}

On the other hand, the largest-order-statistic policy is also stationary. The state transitions of the $i$-th switch are given by the same Markov chain in Fig.~\ref{fig:P2} with
\begin{eqnarray*}
q_i = \frac{i^G-(i-1)^G}{M^G},~~
a_i = (1-q_i)(1-p_i).
\end{eqnarray*}

Thus, we have
\begin{eqnarray*}
\mathbb{E}[n_i] = \frac{a_i}{1-a_i} = \frac{(1-q_i)(1-p_i)}{1-(1-q_i)(1-p_i)},
\end{eqnarray*}
and
\begin{eqnarray}
J_\textup{order} = \sum_{i=1}^{M} \varphi_i \mathbb{E}[n_i] = \sum_{i=1}^{M} \frac{\varphi_i(1-q_i)(1-p_i)}{1-(1-q_i)(1-p_i)}.
\end{eqnarray}

Let $M\rightarrow\infty$, and $p_1=p_2=\cdots=p_M=p$, we have $q_i\rightarrow 0$, and
\begin{eqnarray*}
\lim_{M\rightarrow\infty}J_\textup{order} = \frac{1-p}{(1-\sigma)p}.
\end{eqnarray*}

\section{Performance of the weighted-probability sampling policy}\label{sec:AppC}
A weighted-probability policy samples the $M$ switches in each slot following the same distribution $\{w_i:i=1,2,...,M\}$. To derive the minimal average cost achieved by the weighted-probability policy, we have to find the optimal distribution $\{w^*_i:i=1,2,...,M\}$.

When operated with the weighted-probability policy, the state transitions of a single switch are the shown in Fig.~\ref{fig:P2}, where $a_i$ is the probability that the i-th switch is not sampled in a slot, giving
\begin{eqnarray*}
a_i = (1-w_i)(1-p_i).
\end{eqnarray*}

Similar to \eqref{eq:B_3} and \eqref{eq:B_4}, we can compute the steady-state distribution of all states in equilibrium, and write the average cost achieved by the weighted-probability policy with distribution $\{w_i:i=1,2,...,M\}$ as
\begin{eqnarray}
J_\textup{weighted} = \sum_{i=1}^{M} \varphi_i \frac{a_i}{1-a_i}
= \sum_{i=1}^{M} \frac{\varphi_i(1-w_i)(1-p_i)}{1-(1-w_i)(1-p_i)},
\end{eqnarray}
and the minimal average cost
\begin{eqnarray}\label{eq:D_1}
&&\hspace{-0.2cm} J^*_\textup{weighted} = \min_{w_i}J_\textup{weighted}
= \min_{w_i}\sum_{i=1}^{M} \frac{\varphi_i(1-w_i)(1-p_i)}{1-(1-w_i)(1-p_i)}   \nonumber\\
&&\hspace{-0.2cm} s.t.  \sum_{i=1}^M w_i = 1,~~w_i \geq 0.
\end{eqnarray}

The optimal solution to this linear program is the same as \eqref{eq:A_7}, i.e., the optimal distribution is
\begin{eqnarray}
w^*_i = \left(v\sqrt{\frac{\varphi_i}{1-p_i}} -\frac{p_i}{1-p_i}  \right)^+,
\end{eqnarray}
and $v$ is chosen such that
\begin{eqnarray*}
\sum_{i=1}^{M} \left(v\sqrt{\frac{\varphi_i}{1-p_i}} -\frac{p_i}{1-p_i}  \right)^+ = 1.
\end{eqnarray*}

The minimal average cost achieved by the weighted-prob\-ability policy is thus twice the lower bound in Theorem \ref{thm:1}, giving
\begin{eqnarray*}
J^*_\textup{weighted} =  \sum_{i=1}^{M}\varphi_i\left(\frac{1}{(1-p_i)w^*_i+p_i}-1\right).
\end{eqnarray*}

%

\section{Proof of Indexability}\label{sec:AppE}
This appendix proves Proposition~\ref{thm:5}.

\begin{NewProof}
We first assume the optimal policy to the decoupled problem has a threshold structure, and derive the relationship between the threshold and the gain of the MDP. Then, we verify that the relationship satisfies the Bellman equation in \eqref{eq:V_bellman2}, hence the threshold policy is the optimal policy to the decoupled problem.

Let us assume there exists a threshold $\Gamma$ such that: if a state $n<\Gamma$, the optimal policy at this state is ``rest'', and 2) if a state $n\geq\Gamma$, the optimal policy at this state is ``sample''.

Without loss of generality, let us set $\bm{h}^*[0] = 0$. From \eqref{eq:V_bellman2}, we have
\begin{eqnarray}
\label{eq:V_F1}
&&\hspace{-0.2cm} c \geq (1-p) \bm{h}^*[n+1], ~~ \forall~n<\Gamma, \\
\label{eq:V_F2}
&&\hspace{-0.2cm} \bm{h}^*[n] =\varphi n + (1-p) \bm{h}^*[n+1] - g^*, ~~ \forall~n<\Gamma,
\end{eqnarray}

For $n\geq\Gamma$, we have
\begin{eqnarray}\label{eq:V_F4}
c \leq (1-p) \bm{h}^*[n+1], ~~ \forall~n\geq\Gamma,
\end{eqnarray}
\begin{eqnarray}\label{eq:V_F5}
\bm{h}^*[n] =c+\varphi n - g^*, ~~ \forall~n\geq\Gamma,
\end{eqnarray}

Next, we show that $\bm{h}^*[n]$ is a monotonically increasing function of $n$ if the optimal policy is a threshold policy.

1) For $n\geq\Gamma$, \eqref{eq:V_F5} indicates that $\bm{h}^*[n]$ monotonically increases with the increase of $n$.

2) At the threshold,
\begin{eqnarray}\label{eq:V_F6}
\bm{h}^*[\Gamma]\!-\!\bm{h}^*[\Gamma\!-\!1] \!\!\!\!&=&\!\!\!\! \bm{h}^*[\Gamma] \!-\! \varphi (\Gamma \!-\! 1) \!-\! (1\!-\!p) \bm{h}^*[\Gamma] + g^* \nonumber\\
\!\!\!\!&=&\!\!\!\! p(c+\varphi\Gamma-g^*) - \varphi(\Gamma - 1) + g^* \nonumber\\
\!\!\!\!&=&\!\!\!\! (1-p)g^* + pc + \varphi - \varphi\Gamma (1-p).
\end{eqnarray}

Let $n=\Gamma - 1$ in \eqref{eq:V_F1},
\begin{eqnarray*}
c\geq (1-p) \bm{h}^*[\Gamma] = (1-p)(c+\varphi\Gamma - g^*)
\end{eqnarray*}
\begin{eqnarray}\label{eq:V_F7}
(1-p)g^* +pc \geq \varphi\Gamma (1-p)
\end{eqnarray}

Substituting \eqref{eq:V_F7} into \eqref{eq:V_F6} gives us $\bm{h}^*[\Gamma]-\bm{h}^*[\Gamma-1] > 0$.

3) For $n<\Gamma$, it follows from \eqref{eq:V_F2} that
\begin{eqnarray*}
\bm{h}^*[n]-\bm{h}^*[n-1] = \varphi + (1-p)(\bm{h}^*[n+1]-\bm{h}^*[n]).
\end{eqnarray*}

Since $\bm{h}^*[\Gamma]-\bm{h}^*[\Gamma-1] > 0$, we have
\begin{eqnarray*}
\bm{h}^*[n]-\bm{h}^*[n-1] > 0,  ~~ \forall~n<\Gamma.
\end{eqnarray*}
Overall, $\bm{h}^*[n]$ is a monotonically increasing function of $n$.

Next, we show that the threshold policy satisfies the Bellman equation in \eqref{eq:V_bellman2}, and hence is the optimal policy to the decoupled problem.

Consider any state $n$. If the optimal action at state $n$ is ``sample'', we must have $n\geq\Gamma$ for the threshold policy. Thus, $\bm{h}^*[n+1]\geq\bm{h}^*[\Gamma+1]$ because $\bm{h}^*[n]$ is a monotonically increasing function of $n$. Then,
$(1-p)\bm{h}^*[n+1] \geq (1-p)\bm{h}^*[\Gamma+1] \geq c$.
This is consistent with \eqref{eq:V_bellman2} if the optimal action at state $n$ is ``sample''.

On the other hand, if the optimal action at state n is ``rest'', we must have $n\leq\Gamma-1$ for the threshold policy, and $\bm{h}^*[n+1]\leq\bm{h}^*[\Gamma]$. Thus,
\begin{eqnarray*}
(1-p)\bm{h}^*[n+1] \leq (1-p)\bm{h}^*[\Gamma] \leq c.
\end{eqnarray*}
This is consistent with \eqref{eq:V_bellman2} if the optimal action at state $n$ is ``rest''.

In conclusion, the optimal policy for the decoupled problem is a threshold policy.
\end{NewProof}

\section{}\label{sec:AppF}
This appendix proves Theorem~\ref{thm:7}.

\begin{NewProof}
The Whittle index at state $n$ is the sampling cost $c^*$ for which ``sample'' and ``rest'' make no difference. As suggested by the Bellman equation in \eqref{eq:V_bellman2}, we have
\begin{eqnarray}\label{eq:V_F15}
c^* = (1-p) \bm{h}^*[n+1],
\end{eqnarray}
That is, if $c<c^*$, the optimal policy at the state $n$ is to sample, if $c>c^*$, the optimal policy at the state $n$ is to rest. When the sampling cost is exactly $c^*$, it is equally optimal to ``sample'' and ``rest'', and state $n$ is the threshold.

Substituting \eqref{eq:V_F15} into \eqref{eq:V_F5} gives us
\begin{eqnarray}\label{eq:V_F16}
p c^* = (1-p) \varphi (n+1) - (1-p) g^*,
\end{eqnarray}
Next, we derive $g^*$ as a function of $c^*$.

Let $n=0,1,2,...,n-1$ in \eqref{eq:V_F2}, we have
\begin{small}
\begin{eqnarray}
&&\hspace{-0.6cm}\bm{h}^*[1] = g^* \frac{1}{1-p}, \nonumber\\
&&\hspace{-0.6cm}\bm{h}^*[2] = g^*\left(\frac{1}{1-p} + \frac{1}{(1-p)^2} \right) - \varphi \left(\frac{1}{1-p} \right), \nonumber\\
&&\hspace{-0.6cm}\bm{h}^*[3] \!=\! g^*\!\!\left(\!\frac{1}{1\!-\!p}\! +\! \frac{1}{(1\!-\!p)^2} \!+\! \frac{1}{(1\!-\!p)^3} \!\right) \!-\! \varphi\! \left(\!\frac{2}{1\!-\!p} \!+\! \frac{1}{(1\!-\!p)^2} \!\right), \nonumber\\
&&\hspace{-0.6cm}\cdots \nonumber\\
\label{eq:V_F17}
&&\hspace{-0.6cm}\bm{h}^*[n] = g^*\left(\frac{1}{1-p} + \frac{1}{(1-p)^2} + \cdots + \frac{1}{(1-p)^n} \right) - \nonumber\\
&&\hspace{0.5cm}  \varphi \left(\frac{n-1}{1-p} + \frac{n-2}{(1-p)^2}+ \cdots+ \frac{1}{(1-p)^{n-1}} \right) \nonumber\\
&&\hspace{0.1cm}= g^*\frac{(1-p)^{-n}\!-\!1}{p} - \varphi \frac{(1-p)^{1-n}\!-\!(1-p)\!-\!np}{p^2}.
\end{eqnarray}
\end{small}

On the other hand, since $n$ is the threshold, we have \eqref{eq:V_F18} from \eqref{eq:V_F5}.
\begin{eqnarray}\label{eq:V_F18}
\bm{h}^*[n] = c^* +\varphi n - g^*.
\end{eqnarray}

Equating \eqref{eq:V_F17} and \eqref{eq:V_F18} gives us
\begin{eqnarray}\label{eq:V_F19}
g^* = \frac{p(1\!-\!p)^n}{1\!-\!(1\!-\!p)^{n+1}}c \!+\! \varphi n \!-\! \frac{(1\!-\!p)^{n+1}\!-\!(1\!-\!p)+np}{p\left[1 \!-\! (1\!-\!p)^{n+1} \right]}\varphi
\end{eqnarray}

Substituting \eqref{eq:V_F19} into \eqref{eq:V_F16}, we finally have
\begin{eqnarray*}
c^* = \frac{\varphi(1-p)}{p^2}\left[ (1-p)^{n+2} + (n+2)p - 1 \right].
\end{eqnarray*}

This is the cost that the controller is willing to pay to sample a device when it is in state $n$. At a decision epoch, the controller computes a $c^*(n_i)$ for each device based on its current state and samples the one with the greatest $c^*(n_i)$.
\end{NewProof}

\section{}\label{sec:AppG}
This appendix evaluates the Bernoulli assumption in Section \ref{sec:II} via simulations.
We focus on one IoT device in the network and suppose the considered IoT device is a crosspoint of $K$ flow paths. When operated with different sampling policies, we shall verify that the inter-sampling time of the crosspoint (measured by the number of slots) has a geometric probability mass function (PMF). In other words, let the inter-sampling time be a discrete random variable $Z$, then
\begin{eqnarray}\label{eq:r1}
\text{Pr}(Z=z) = (1-p)^{z-1}p.
\end{eqnarray}
If \eqref{eq:r1} holds, the event that the IoT device is sampled in a slot (i.e., $H_i^t$, as defined in the manuscript) follows independent Bernoulli distribution with parameter $p$. In the following, we let the controller sample the $K$ flow paths independently using various policies devised in this paper, and verify if \eqref{eq:r1} holds.

{\bf Simulation setup.} Suppose there are $K = 20$ flow paths crossing the considered crosspoint. The number of IoT devices on each flow path, i.e., $M_k\in\mathcal{Z}$, is generated uniformly from $[3, 200]$. The location of the crosspoint on each path is uniformly sampled from $[1, M_k]$.

{\bf State-independent policies.} First, we let the controller sample the $K$ flow paths independently using state-independent policies. As discussed in Section IV of our paper, state-independent policies sample the devices on a flow path following some stationary probability distributions, and the sampling actions over consecutive slots are independent. Thus, we can directly show that the Bernoulli assumption holds for state-independent policies. Take the uniform sampling policy for example. With the uniform sampling policy, the probability that the crosspoint is sampled in a slot is given by
\begin{eqnarray}\label{eq:r2}
p=1-\prod_{k=1}^{K} \left(1-\frac{1}{M_k}\right).
\end{eqnarray}
Thus, $H_i^t$ follows independent Bernoulli distribution with parameter $p$ given in \eqref{eq:r2}.

\begin{figure*}[t]
\centering
\subfigure{
\includegraphics[width=0.35\linewidth]{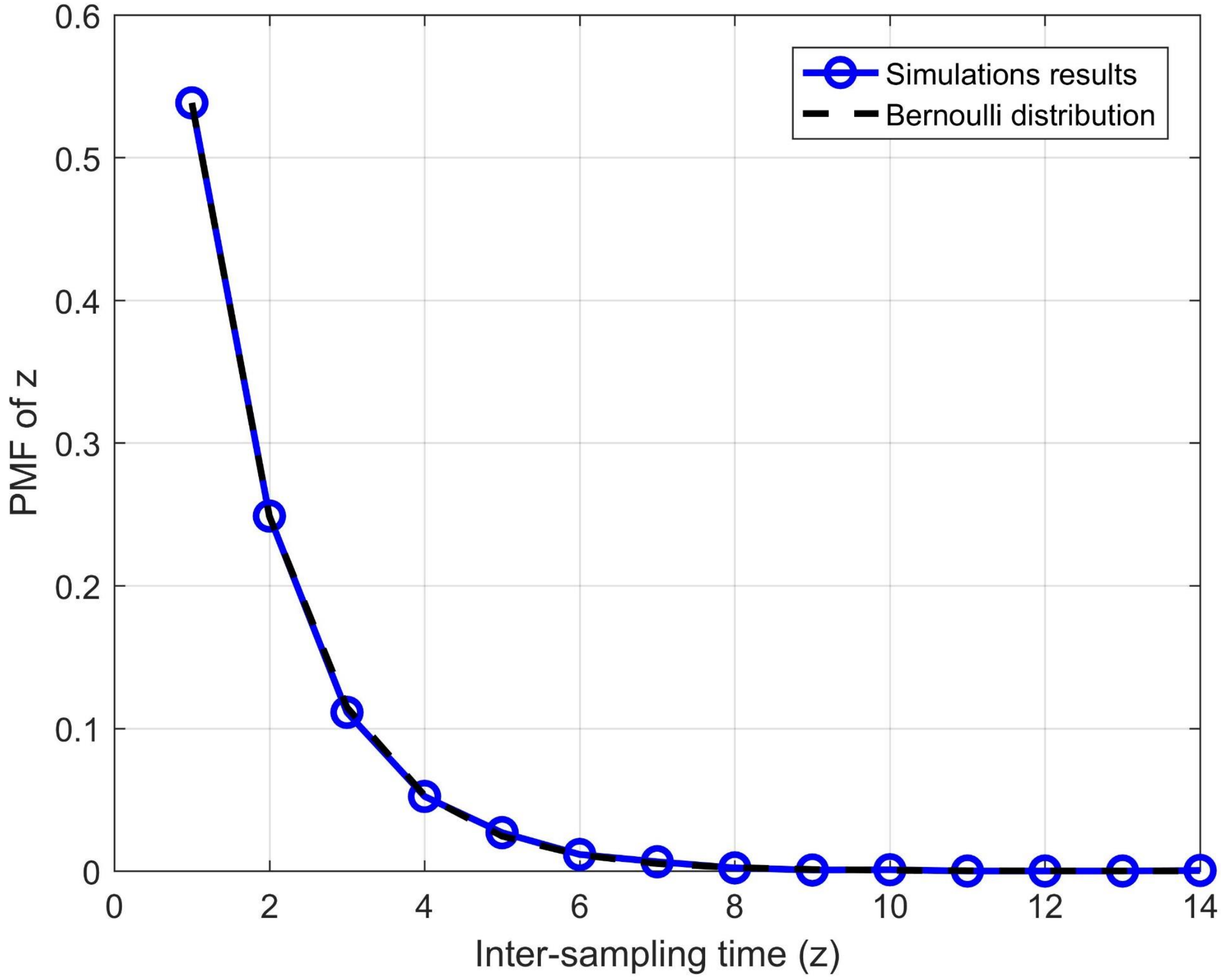}
}
\quad
\subfigure{
\includegraphics[width=0.35\linewidth]{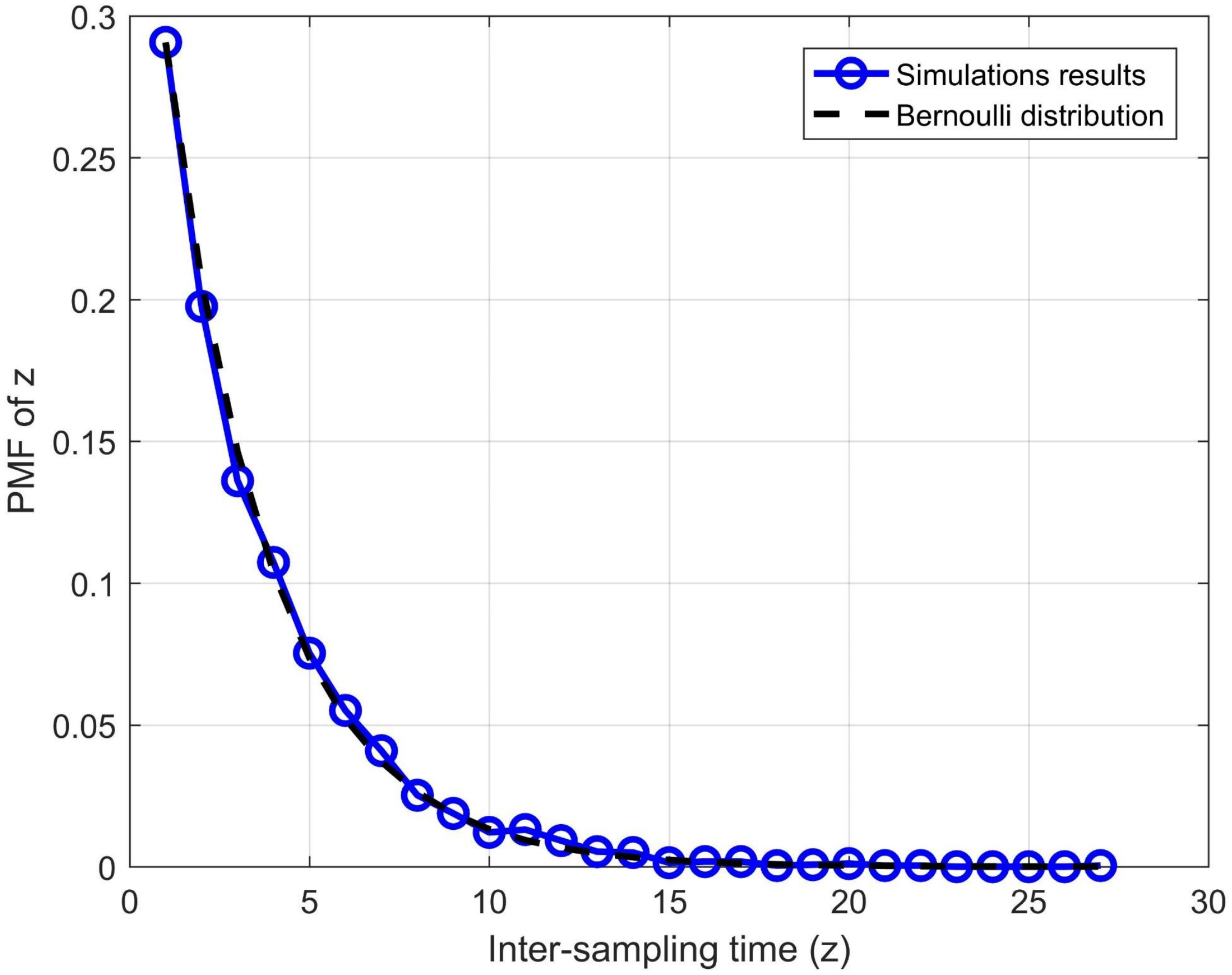}
}
\caption{PMF of the inter-sampling time for the considered IoT device (crosspoint). The controller samples the $K$ flow paths independently using the uniform sampling policy (left) and the la\-rge\-st-or\-der-sta\-tistic policy (right, $G = 2$).}
\label{fig:r4}
\end{figure*}

Simulation results also verify that the Bernoulli assumption holds for state-independent policies. Fig.~\ref{fig:r4} presents the PMF of the inter-sampling time for the considered crosspoint, with the uniform sampling policy (left) and the larg\-est-ord\-er-sta\-tistic sampling policy (right). To obtain Fig.~\ref{fig:r4}, we first run simulations to obtain the PMF of the inter-sampling time (labeled as ``simulation results''). Then, we artificially generate a Bernoulli distribution, which is parameterized by $p=\text{Pr}(Z=1)$ from the simulation results, and plot its PMF on the same figure (labeled as ``Bernoulli distribution'').

As can be seen, for both uniform and largest-order-statistic sampling policies, the PMFs of the inter-sampling time for the considered IoT device fit the corresponding Bernoulli distributions very well.

\begin{figure*}[t]
\centering
\subfigure{
\includegraphics[width=0.35\linewidth]{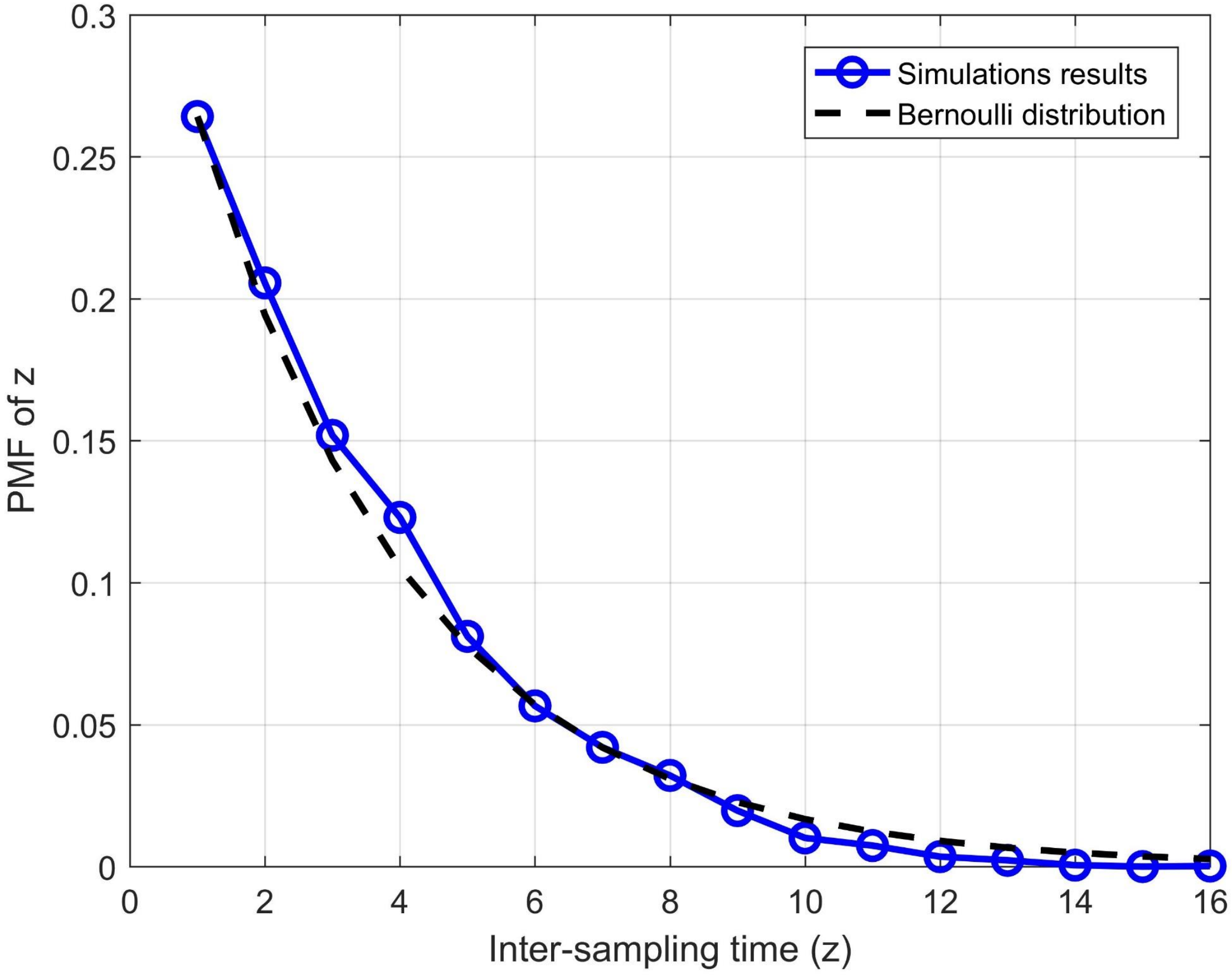}
}
\quad
\subfigure{
\includegraphics[width=0.35\linewidth]{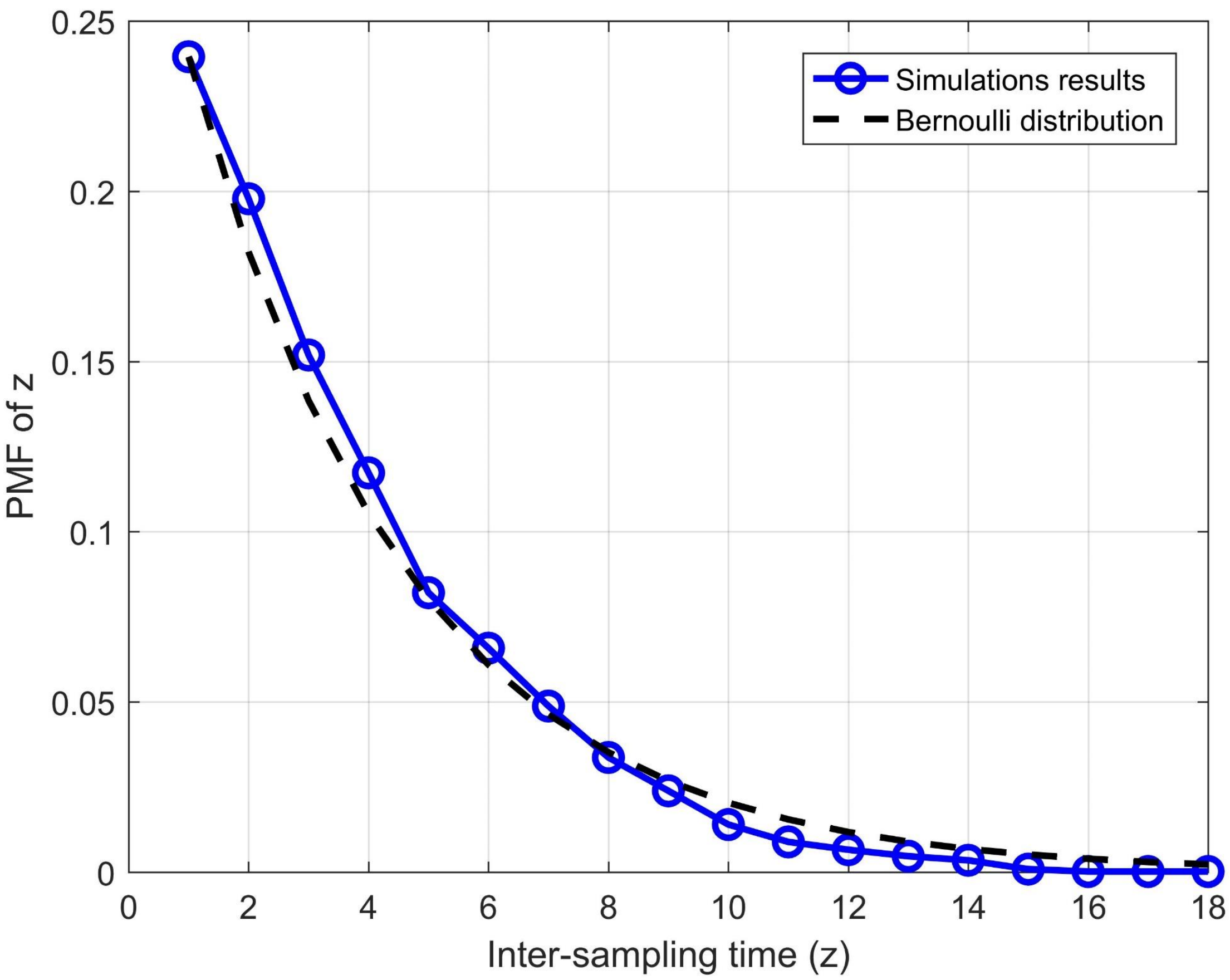}
}
\caption{PMF of the inter-sampling time for the considered IoT device (crosspoint). The controller samples the $K$ flow paths independently using the Whittle index policy (left) and the second-order index policy (right). When using the Whittle index policy, the controller computes the Whittle indexes for all the devices on the $k$-th path by setting $p_i=0.3$ and $\varphi_i=\sigma^{M_k-i}$, where $\sigma=0.9$.}
\label{fig:r5}
\end{figure*}

{\bf Index policies.} We next let the controller sample the $K$ flow paths using the index policies devised in this paper, i.e., the Whittle index policy and the second-order index policy. For the considered crosspoint, the PMFs of the inter-sampling time are presented in Fig.~\ref{fig:r5}, wherein a Bernoulli distribution parameterized by $p=\text{Pr}(Z=1)$ is also plotted. As shown, the simulated PMFs fit the corresponding Bernoulli distributions very well with the index policies.

Overall, we conclude that the Bernoulli assumption of $H_i^t$ is a valid assumption with various policies devised in this paper.

\section{}\label{sec:AppH}
This appendix presents additional simulation results to complement Section \ref{sec:VI}.

\textbf{Simulation 1}: The Whittle index policy versus the optimal policy.
In the first simulation, we further evaluate the optimality of the Whittle index policy benchmarked against the optimal policy. As in Fig.~\ref{fig:S1}, we consider a short flow path with $M = 3$ devices.  Fig.~\ref{fig:S1} considers a homogeneous network where the devices have the same $p_i$. In this simulation, we consider a heterogeneous network where the devices have different $p_i$. In particular, we assume $p_i$, $i=1,2,3$ are sampled uniformly from $(0,p]$, i.e., $p_i\sim U(0,p)$, and $p$ is a variable in the simulation.

\begin{figure}[t]
  \centering
  \includegraphics[width=0.75\columnwidth]{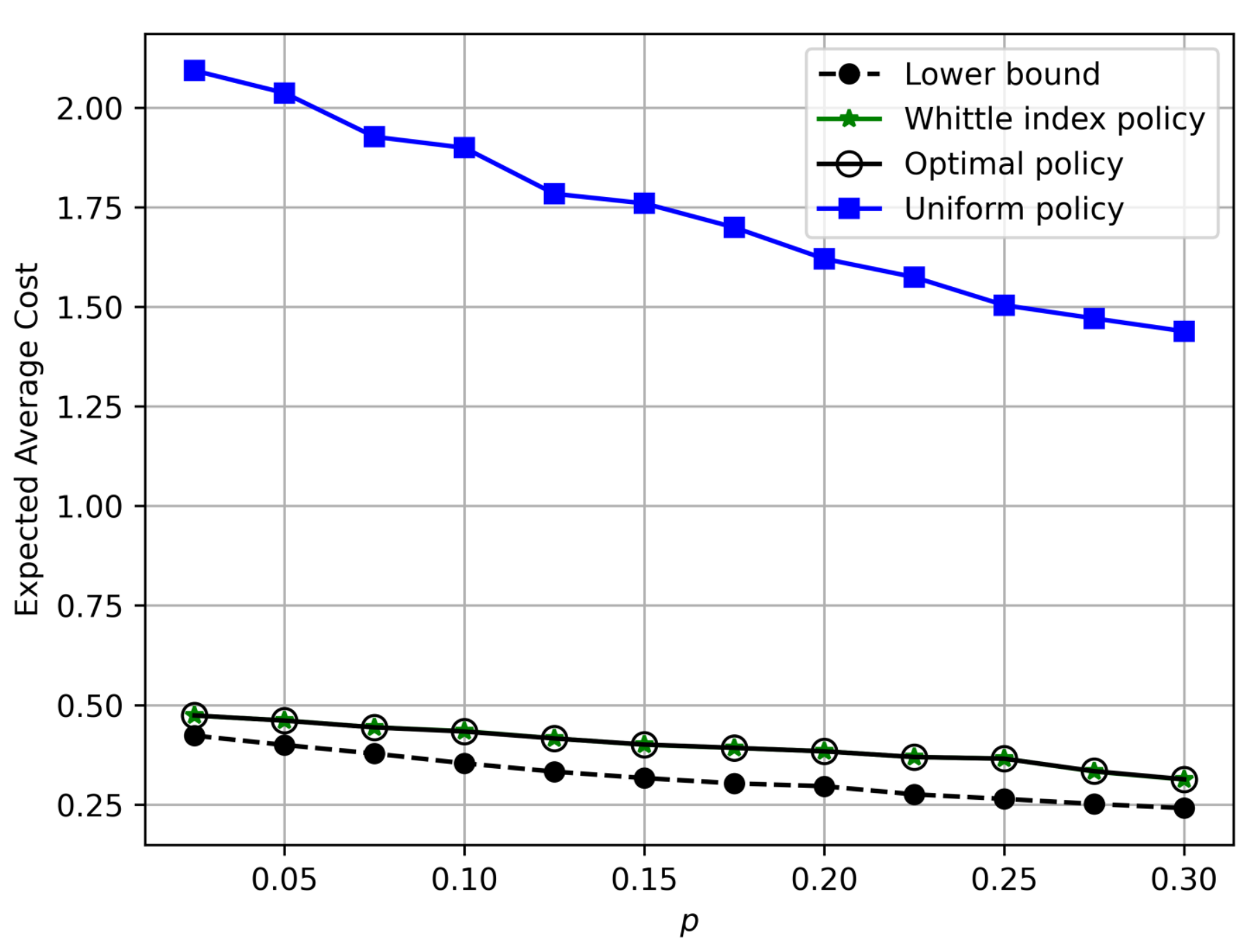}\\
  \caption{Performance comparison between the Whittle index policy and the optimal policy, where $\sigma=0.1$, $M=3$ and $p_i\sim U(0,p)$.}
\label{fig:R1}
\end{figure}

\begin{figure}[t]
  \centering
  \includegraphics[width=0.75\columnwidth]{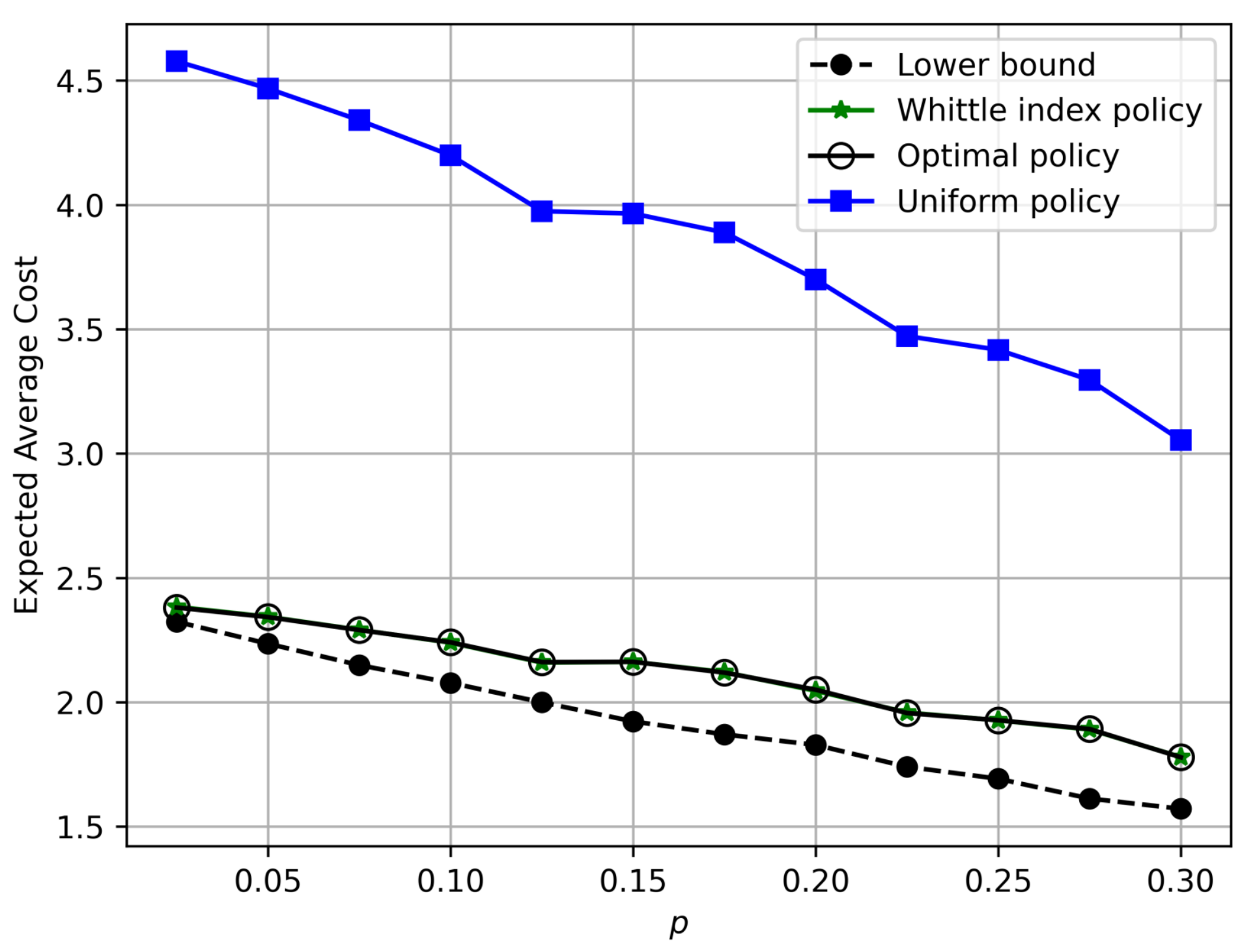}\\
  \caption{Performance comparison between the Whittle index policy and the optimal policy, where $\sigma=0.8$, $M=3$ and $p_i\sim U(0,p)$.}
\label{fig:R2}
\end{figure}

Given the uniformly sampled $p_i$, we simulate the expected average cost achieved by the Whittle index policy and the optimal policy in Fig.~\ref{fig:R1} ($\sigma=0.1$) and Fig.~\ref{fig:R2} ($\sigma=0.8$), respectively. The expected average costs of the uniform sampling policy and the lower bound are also presented as benchmarks. As can be seen, we have the same observations as Fig.~\ref{fig:S1}, the Whittle index policy performs as well as the optimal policy.

\begin{figure}[t]
  \centering
  \includegraphics[width=0.75\columnwidth]{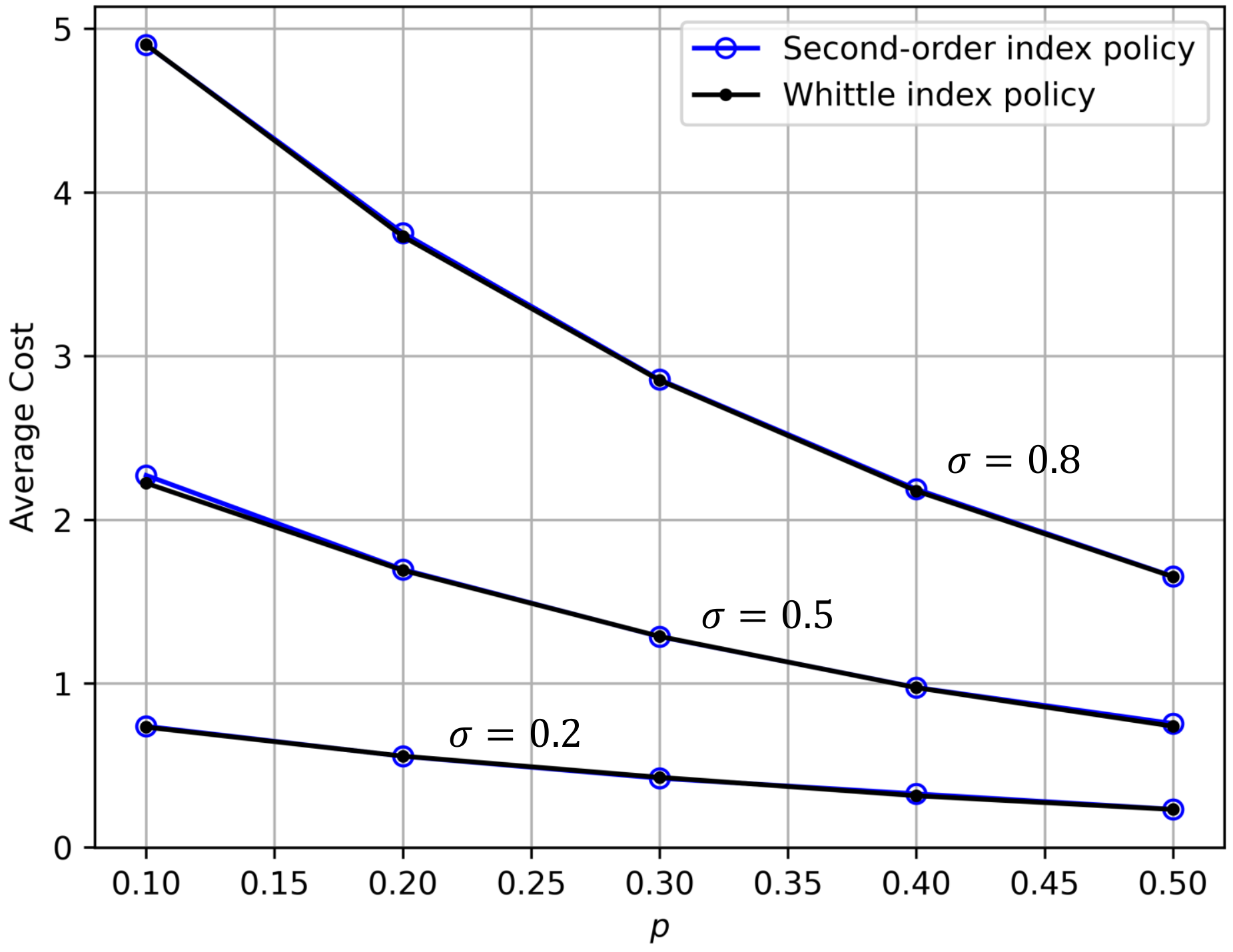}\\
  \caption{Performance comparison between the second-order index policy and the Whittle index policy, where $M=5$ and $\sigma=0.2$, $0.5$, and $0.8$.}
\label{fig:R3}
\end{figure}

\begin{figure}[t]
  \centering
  \includegraphics[width=0.75\columnwidth]{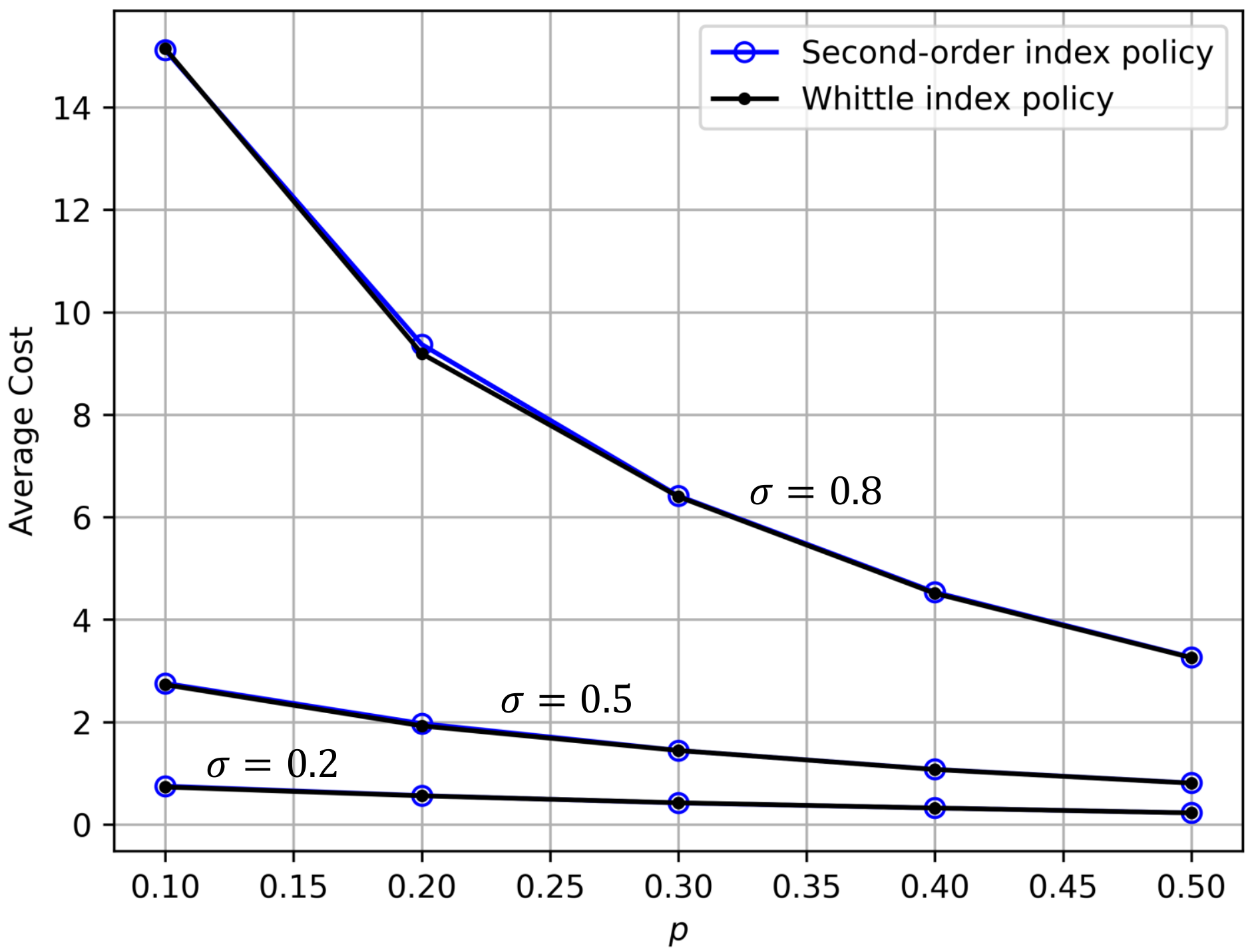}\\
  \caption{Performance comparison between the second-order index policy and the Whittle index policy, where $M=40$ and $\sigma=0.2$, $0.5$, and $0.8$.}
\label{fig:R4}
\end{figure}

\textbf{Simulation 2}: The second-order index policy versus the Whittle index policy.
In the second simulation, we evaluate the performance of the second-order index policy benchmarked against the Whittle index policy. In Fig.~\ref{fig:S3}, we compare the performance of the two policies in a heterogeneous network. It is shown that the performance gaps between the two policies are negligible. In this simulation, we further evaluate the two policies in a homogeneous network where $p_1=p_2=...=p_M=p$, $M=5$, $40$, and $\sigma=0.2$, $0.5$, and $0.8$.
The simulation results in Fig.~\ref{fig:R3} and \ref{fig:R4} confirm that the performances of the second-order index policy and the Whittle index policy are on the equal footing in terms of the average cost performance.

In summary, we have shown that the two main results in our paper are robust to various parameter setting.
On the one hand, the Whittle index policy is as good as the optimal policy for small $M$;
on the other hand, the second-order index policy is on the equal footing with the Whittle index policy while requiring no network prior information to be executed.

\bibliographystyle{IEEEtran}
\bibliography{References}

\end{document}